\documentclass[letterpaper,twocolumn,10pt]{article}
\usepackage{usenix-2020-09}

\newif\ifdraft
\drafttrue
\newif\ifcomment
\commenttrue

\draftfalse

\ifdraft
  \makeatletter
  \def\ps@mydraftstyle{%
    \def\@oddhead{\color{red}{DRAFT \hfil \@date \hfil Strictly Confidential}}\relax
    \def\@evenhead{\color{red}{DRAFT \hfil \@date \hfil Strictly Confidential}}\relax
    \def\@oddfoot{\hfil\thepage\hfil}
    \def\@evenfoot{\hfil\thepage\hfil}
  }
  \makeatother
  \let\OldMaketitle\maketitle
  \renewcommand{\maketitle}{\OldMaketitle\thispagestyle{mydraftstyle}}
\fi
\ifcomment
  \def\comment#1{{\color{magenta}{\small\em #1 }}}
\else
  \def\comment#1{{}}
\fi

\usepackage{tikz}
\usepackage{amsmath}
\usepackage{cite}
\usepackage{amsmath,amssymb,amsfonts}
\usepackage{algorithm}
\usepackage{algorithmic}

\begin{document}

\date{\today}

\title{\Large \bf Towards End-to-End Error Management for a Quantum Internet}

\author{
{\rm Shota Nagayama}\\
Mercari, Inc., 
Roppongi Hills Mori Tower 18F,
6-10-1 Roppongi, Minato-ku, 
Tokyo, 106-6118, Japan
} 

\maketitle

\begin{abstract}
Error management in the quantum Internet requires stateful and stochastic processing across multiple nodes, which is a significant burden.
In view of the history of the current Internet, the end-to-end principle was devised for error management, simplifying the work inside the network and contributing significantly to the scalability of the Internet.  
In this paper, we propose to bring the end-to-end principle into the error management of quantum Internet to improve the communication resource utilization efficiency of a quantum Internet.
The simulation results show that the error management using the end-to-end principle and locality can be more resource-efficient than other settings.
In addition, when end-to-end error management is used, if the error probability of qubits in the end node is sufficiently low, there is no problem even if the error probability on the network side is higher than that in the end node, and the load on the network can be reduced.
Our proposal will contribute to improving the communication capacity and scalability of the quantum Internet, as well as to improve the interoperability of quantum Autonomous Systems.
In addition, existing studies on routing and other aspects of the quantum Internet may exclude error management from their scope due to its complexity. The results of this study provide validity to the assumptions of such studies.
\end{abstract}

\section{Introduction}
Unlike other next-generation networks such as Beyond 5G, the Quantum Internet is the only one capable of transmitting quantum data and is expected to be used in conjunction with the current Internet to support human society in the future full-scale quantum technology era.

Research on quantum Internet can be roughly divided into three categories: architecture, hardware, and applications. 
In terms of applications, for example, by connecting quantum computers, distributed computation problems such as the Byzantine General Problem can be solved faster~\cite{Ben-Or2005}. Besides distributed algorithms, distributed quantum computation improves the scalability of quantum computation~\cite{Grover1997a,Cirac1999,Steane2000,VanMeter2006,Nagayama2017a}. While the computational power of current computers is doubled at best when two computers are parallelized, the solution space that can be handled becomes exponentially larger when two quantum computers are connected~\cite{Caleffi2018a}. Thus, the concepts of parallelism and scalability are fundamentally different from those of current-generation computers. A global quantum internet could enable us to connect quantum computers around the world to solve particularly large problems. In addition to quantum computation, a variety of other applications have been found for the quantum Internet. For example, the quantum Internet's algorithm for generating a secret shared key differs from current public-key cryptography in that not enough information flows over the Internet to break it~\cite{Bennett1984,Ekert1991}. Therefore, it has the peculiarity of being theoretically unbreakable. In security, in addition to communication confidentiality, there are such as fingerprinting~\cite{Buhrman2001} for authentication and Blind Quantum Computation, which is important for data privacy~\cite{Broadbent2009, Morimae2013}. A variety of applications have also been discovered that are not limited to computational fields, such as ultra-precision time synchronization~\cite{jozsa2000}, ultra-long baseline radio telescopes~\cite{Gottesman2012}, and reference frame correction~\cite{Rudolph2003}. It is expected that more applications will be discovered in the future.

    Signal repeating technology is an important technology in computer networks, and any node in the current Internet performs signal repeating. Hardware research in the quantum Internet has long been hampered by the difficulty of demonstrating the principle of quantum repeating in experimental research. \footnote{
      This difficulty led to a spin-out based on the idea of terminating each quantum communication application at each node without performing quantum signal repeating. This network is called quantum key distribution network. Both are called quantum networks and are easily confused, but only the quantum Internet can run end-to-end quantum communication applications and is a general-purpose quantum communication network.}
      In 2019, quantum memory was used to apply quantum operations to photons arriving at different times ~\cite{Bhaskar2020}. In recent years, proof of principle for quantum signal repeating has been achieved ~\cite{Pompili2021,Pu2021}. A quantum network version of the all-optical architecture, which has been the focus of much attention in classical networks, has also been studied, and a proof of concept for all-optical quantum repeating has been successfully achieved ~\cite{Hasegawa2019,Li2019}.
      Experiments to generate quantum entanglement in a 22 km field environment have also been successfully conducted~\cite{Yu2020}.

    As research on manipulating quanta at will as qubits (quantum bits) continues, the importance of research on the computer networking side, such as routing algorithms, architectures, and other mechanisms for assembling qubits as a large-scale computer network, is increasing. Algorithms and systems for quantum computer networking should be based on the idea that the initial small-scale network will be gradually extended, interconnected, and developed into a large-scale network~\cite{Meter2011}.
    Architectures and simulations for running huge quantum networks, hence quantum Internet, will become more and more important~\cite{van-meter21:aqia, satoh21:_quisp}.

    Among the many issues, the fundamental issues include the generation of a Bell pair, which is a pair of qubits in a unique quantum correlation, at the link level, the generation of end-to-end bell pairs through many intermediate nodes such as quantum routers and quantum repeaters, end-to-end connection management, resource management across domains, and error management~\cite{irtf-qirg-principles-07}. In addition, resource management and error management across each domain can be mentioned first. 
    
    Some proposals use local information to perform routing to generate multi-hop bell pairs~\cite{Chakraborty,Gyongyosi2018a}, proposals that use global information to determine routes~\cite{VanMeter2013,Caleffi2017,Chakraborty2020}, and proposals that reserve network resources over a large area and "connect" successful link Bell pairs to generate end-to-end Bell pairs~\cite{Das2017,Pant2019,Shi2020}. By exploiting quantum characteristics, creating Bell pairs between non-neighboring nodes in advance may reduce quantum operations after an actual request is made, and thus can answer the request quickly~\cite{Schoute}. However, as the number of requests increases, there is no room to create and store non-adjacent Bell pairs in advance, and thus no gain~\cite{Chakraborty}. Such methods are considered suitable for environments not saturated with requests, such as some VPNs. There is also a proposal for a routing method to remotely generate multi-qubit entangled states called graph states~\cite{Pirker2019}. 
    Multiplexing is also shown to be useful for resource management in the quantum Internet, and the effectiveness of time-division multiplexing, space-division multiplexing, and statistical multiplexing has been reported~\cite{Aparicio}. 

    At the link level, there is a protocol that has been tested in elaborate simulations~\cite{Dahlberg2019}. 
    A quantum network protocol designed to enable multi-hop quantum communication has also been proposed~\cite{Kozlowski2020}.
    There is also a proposal for a method to establish end-to-end connections~\cite{Matsuo2019}. 
    
  \subsection{Our contribution}
  Entanglement purification is a practical error management scheme that detects errors of Bell pairs in quantum networks.
  Purification is essential for quantum communication because it is an excellent method to eliminate the effect of noise even in a noisy communication environment, and it is less challenging to implement than other error management schemes.
  However, purification is a complicated and synchronous process between nodes that share Bell pairs, and if it is used frequently, the load of the network gets heavy.
  The reason for this complexity is that the success or failure of purification is probabilistic, and this probability varies depending on the input state and requires communication between remote nodes to confirm the success or failure. The details of purification are described in Subsec.~\ref{subsec:ep}. A load of purification on the network is high because the quantum memory must be maintained until the success or failure can be confirmed. In particular, if it is necessary to perform purification among intermediate nodes, multiple communications may block each other's resources, resulting in a significant loss of scalability.
  In other words, although purification is excellent prior research, its operational guidelines are still unclear when we try to apply it in actual network-wide operations.
  In this study, we introduce the end-to-end principle into the entanglement purification scheme. We hypothesize that error management based on the end-to-end principle ~\cite{Saltzer1984}, which has been influential in the research and development of the current Internet, is also effective in the quantum Internet and verify this hypothesis by simulation. In this context, we divide the purification into the following three patterns and analyze the effects.
  \begin{description}
      \item[link-level purification] This is the purification between neighboring nodes. It can be viewed as part of the generation of link quantum entanglement.
      \item[internetworking-level purification] This is the purification between non-adjacent nodes. Since the purification is performed between distant nodes, the communication time to confirm the success or failure is long, and the load is high.
      \item[end-to-end-level purification] Although the communication time to confirm success or failure is the longest since this is purification between end nodes, it can be considered separately from the network load.
  \end{description}
  In this study, we define "error management that brings in the end-to-end principle" as not performing purification at the internetworking level, which is a hefty load for the network but performing purification only at the local link-level and end-to-end level.

As a result, error management using the end-to-end principle is found to be more resource-efficient than other settings.
The simulations also confirmed that the locality is important in the quantum Internet, just as the locality of data and computational resources through CDNs is very effective in the current Internet.
The end-to-end error management is also helpful in building a quantum Internet with qubits from physical systems such as atomic ensembles and rare-earth elements, where complex quantum operations such as purification are hard to implement.
In addition, existing studies on algorithms such as routing for the quantum Internet often exclude error management from their schemes due to its complexity. The results of this study provide validity to the schemes of such studies. 

Meanwhile, some applications cannot be localized, such as cryptographic communication between locations. If the connection is so long, purification at the internetworking level may be necessary.
It is possible to perform purification at an intermediate node, but it may block other communication that the node handles. To solve this problem, it may be practical to attach a "side box" to the intermediate node for performing long-range purification instead of the intermediate node.
The quantum Interment may consist of the end-to-end protocol for general use and the side box protocol for very long-distance use.
The end-to-end nature of the error management and the two-protocol system proposed in this study will contribute significantly to the scalability of the quantum Internet when it is constructed as a large-scale distributed system.

\subsection{Related Work}
There have been several studies on entanglement purification for a quantum Internet. However, none of them considers the operation as a network.
Van Meter {\it et al.} proposed a system design using purification and investigated scheduling algorithms of purification~\cite{VanMeter2009a}. 
They calculated startup latency and showed that link-level purification should be executed as much as possible unless entanglement swapping gate errors and memory errors are large enough to matter. However, they did not consider memory decoherence during classical communication and the end-to-end principle. 
Brand {\it et al.}'s study aimed to calculate the waiting time and fidelity of the first end-to-end Bell pair generated after the start of the communication process and did not consider the network load, nor did it consider the communication time of the classical communication time for purification~\cite{Brand2020}.
Khatri {\it et al.} calculated the average time to generate the first end-to-end bell pair and the maximum hops to create a Bell pair. However, they only assume that purification is included in the protocol for generating raw Bell pairs at the link level and do not explicitly consider it. Also, multi-hop purification, including end-to-end purification, is not considered at all~\cite{Khatri2019}.
Santra {\it et al.} proposed to efficiently connect bell pairs too far distances by using different types of quantum memories depending on the nesting level (which in practice represents the distance) of the entanglement swapping~\cite{Santra2019}. Their method reduces the load on the quantum memory that generates link bell pairs between neighboring nodes but does not consider purification.
Hartmann {\it et al.} showed quantum memory error limits the distance of quantum communication due to memory decoherence during classical communication~\cite{Hartmann2007}. 
These studies do not fully consider the load on the network when generating end-to-end Bell pairs. To build a scalable quantum Internet, it is necessary to consider the network load due to purification.
%
    %
%
    %

\section{Quantum Information and Quantum Networking}

Qubits in quantum information processing are characterized by superposition and quantum entanglement. A qubit is represented by a vector
\begin{equation}
  \vert \psi \rangle = \alpha \vert 0 \rangle + \beta \vert 1 \rangle 
\end{equation}
where
$\vert 0\rangle = 
\begin{pmatrix}
  1\\
  0
\end{pmatrix}$
and
$\vert 1 \rangle = 
\begin{pmatrix}
  0\\
  1
\end{pmatrix}$,
each of which corresponds to 0 and 1 respectively, and where $\alpha$ and $\beta$ are complex numbers. 
The presence of these coefficients expresses the "superposition" phenomenon of a qubit, handling 0 and 1 simultaneously. 
When measuring a qubit, $\vert \alpha \vert ^2$ and $\vert \beta \vert ^2$ are the probabilities of observing 0 and 1, respectively. 
Quantum entanglement is a feature that appears when two or more qubits are present, such as
\begin{equation}
  \vert \Phi^+ \rangle = \frac{1}{\sqrt{2}}(\vert 0_a0_b\rangle + \vert 1_a1_b\rangle)
\end{equation}.
When coin flipping is performed with two coins, there is no correlation between the probabilities of the two outcomes. However, in the case of qubits, correlations can be made. In this example, when the value of qubit $a$ is 0, qubit $b$ is also 0, and when the value of qubit $a$ is 1, qubit $b$ is also 1. This correlation does not include the situation where qubit $b$ is 1 (0) when qubit $a$ is 0 (1). This counterintuitive correlation is a quantum-specific correlation. This correlation state is called quantum entanglement. Since it is a correlation, quantum entanglement has no orientation, though entangled qubits are often delivered from a node to another node. Generating $\vert \Phi^+\rangle $ between any two nodes is the most basic function of the quantum internet~\cite{VanMeter2014}.

\subsection{Quantum Error}
Errors occur in qubits due to noise interaction with the environment and other factors.
Errors can be regarded as the occurrence of unintended gate operations.
For example, an unintentional bit flip operation is a bit flip error (X error), and an unintentional phase flip operation is a phase flip error (Z error).
If a bit flip error, phase flip error, or both occur in either qubit in $\vert \Phi^+ \rangle$, the state changes to
\begin{equation}
  \vert \Psi^+ \rangle = \frac{1}{\sqrt{2}}(\vert 01 \rangle + \vert 10 \rangle)
\end{equation}
\begin{equation}
  \vert \Phi^- \rangle = \frac{1}{\sqrt{2}}(\vert 00 \rangle - \vert 11 \rangle)
\end{equation}
\begin{equation}
  \vert \Psi^- \rangle = \frac{1}{\sqrt{2}}(\vert 01 \rangle - \vert 10 \rangle),
\end{equation} respectively.
If the same error occurs again in either qubit, the state returns to $\vert \Phi^+ \rangle$.
Among maximally entangled two-qubit states, the four states $\vert \Phi^+\rangle$, $\vert \Psi^+\rangle$, $\vert \Phi^-\rangle$, and $\vert \Psi^-\rangle$ are called the Bell states.
In this work, we treat $\vert\Phi^+\rangle$ as the target Bell pair and $\vert\Psi^+\rangle$, $\vert\Phi^-\rangle$, and $\vert\Psi^-\rangle$ as the error states.

A quantum state has an expectation value called fidelity, which measures how close it is to the state it should be in. When fidelity $= 1$, the qubit is in its ideal state. In an error-prone environment, fidelity $< 1$. Therefore, the fidelity of a raw Bell pair generated by a link is lower than 1.
By using a density matrix $\vert\Phi^+\rangle\langle\Phi^+\vert$ and treating it statistically, it is possible to add up the ideal and erroneous states.
Equ.~\ref{equ:state} is the model of the quantum state of a Bell pair which may involve errors.
\begin{equation}
  \rho = A \vert \Phi^+ \rangle \langle \Phi^+ \vert +
         B \vert \Psi^+ \rangle \langle \Psi^+ \vert + 
         C \vert \psi^- \rangle \langle \psi^- \vert + 
         D \vert \Phi^- \rangle \langle \Phi^- \vert 
  \label{equ:state}
\end{equation}
$A$ is the probability of taking the ideal state, which corresponds to fidelity. The $B, C, and D$ are the probabilities of taking each of the states generated by the error.
Since perfect quantum manipulation and quantum memory have not yet been achieved, the fidelity of the quantum state deteriorates with each manipulation and while waiting in memory.

\subsection{Photon Loss and no-cloning theorem}
The quantum Internet uses photons to communicate.
However, sending photons to distant locations can be difficult.
For example, photon loss occurs when photons are absorbed by the optical fiber itself while they are passing through the optical fiber~\cite{Devitt2009}.
Quantitatively, the number of photons arriving depends on the attenuation rate of the optical fiber, which is about $\frac{1}{e}$ every 22 km.
In other words, for a distance L(km)
\begin{equation}
  loss\ rate = \frac{1}{e^{\frac{L}{22}}}
\end{equation}
Furthermore, there is a theorem called the no-cloning theorem, which states that we cannot create a backup for a qubit~\cite{Wootters1982}.
Therefore, passing important quantum data directly through lossy quantum channels, such as optical fibers, is hazardous.

\subsection{Reproducable $\vert \Phi^+ \rangle$ and Quantum Teleportation}
\begin{figure}[t]
  \begin{center}
    \includegraphics[width=8cm]{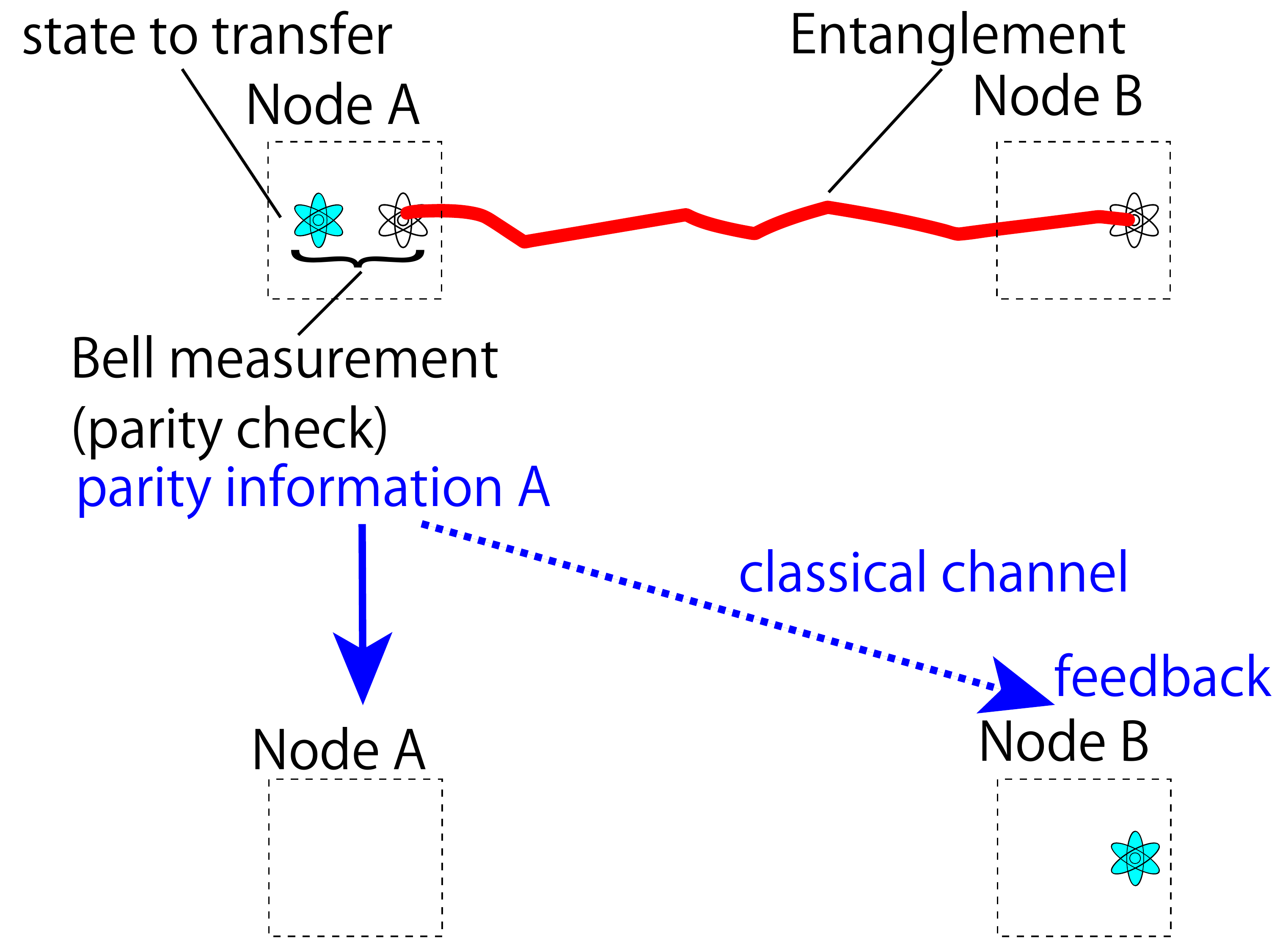}
    \caption{Procedure of quantum teleportation.}
    \label{fig:qt}
  \end{center}
\end{figure}

Quantum teleportation is a method of transmitting quantum data using a lossy quantum channel~\cite{Bennett1993}.
Quantum teleportation makes it possible to transmit an arbitrary qubit to a distant location without using a new quantum channel by consuming a $\vert Phi^+ \rangle $ that already exists.
Fig. ~\ref{fig:qt} shows a diagram of quantum teleportation.
The measurement annihilates the qubit in Node A, but the remaining qubit in Node B becomes the same as the desired qubit.  
Since $\vert \Phi^+ \rangle$ is a resource state that can be created any number of times, it is safe to create it through the lossy channel.
Therefore, in the quantum Internet, we first create a $\vert\Phi^+\rangle$ and then consume the $\vert\Phi^+\rangle$ to perform quantum teleportation to transmit the data we want to send.

\subsection{Entanglement Swapping}
\begin{figure}[t].
\begin{center}
    \includegraphics[width=8cm]{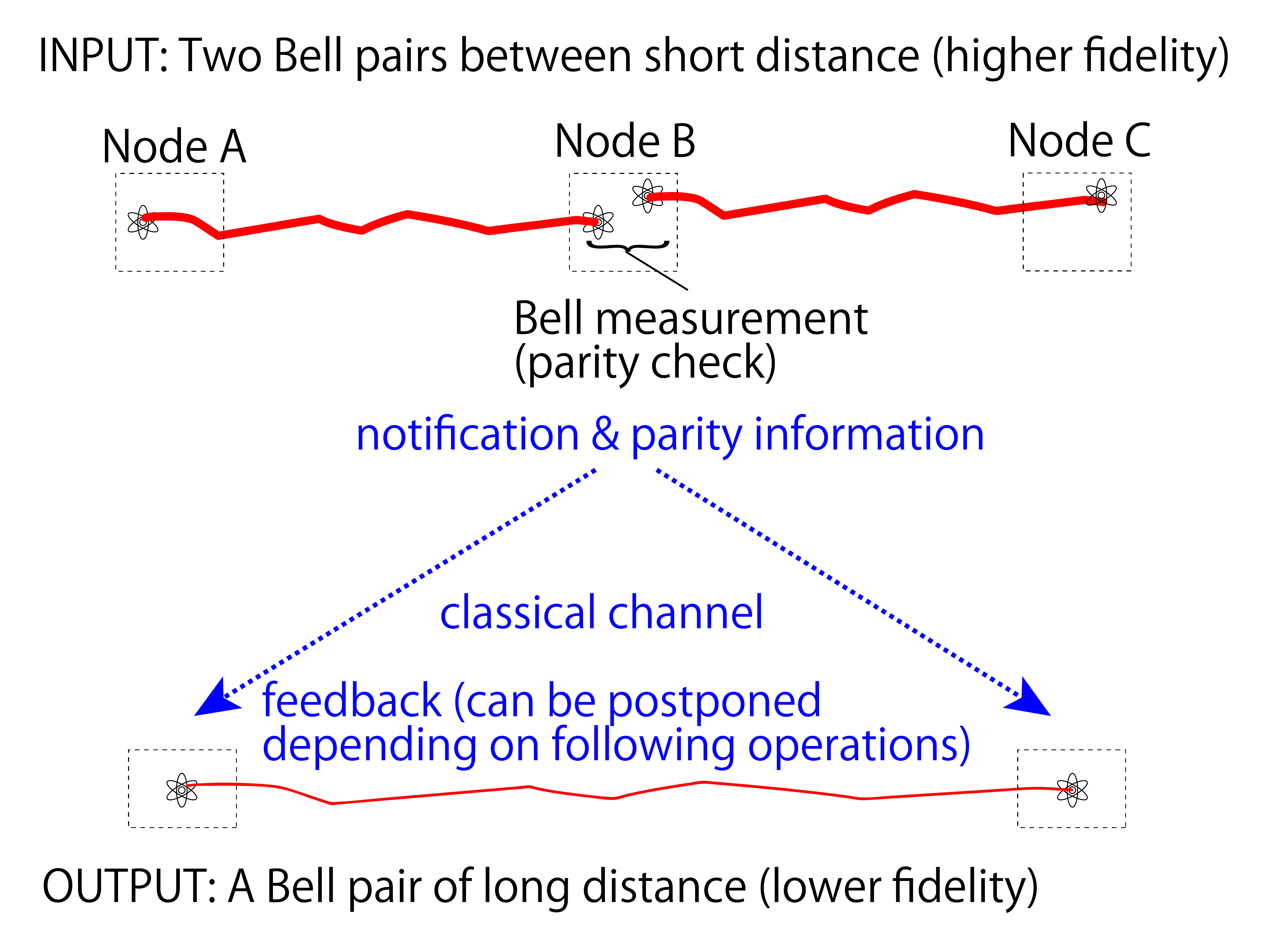}
    \caption{Procedure of an entanglement swapping. 
    Feedback to the residual qubits (X, Y, or Z gate depending on the measurement outcome) may be postponed depending on the following operation.
    Following entanglement swapping and entanglement purification can start even before the feedback.}
    \label{fig:es}
\end{center}
\end{figure}

However, there is still a problem with photon loss.
Since the number of photons that survive from photon absorption decreases exponentially, the direct transmission does not scale with distance, making quantum communication with distant locations impossible.
The solution to this problem is entanglement swapping, which "connects" two Bell pairs to achieve quantum repeating capabilities.
Fig. ~\ref{fig:es} illustrates the entanglement swapping.
We solve the problem of exponentially decreasing reachability by generating Bell pairs in $n$ intervals and "connecting" them with entanglement swapping to generate end-to-end Bell pairs.
This scheme results in a loss probability of
\begin{equation}
  loss\ rate = \frac{1}{e^{\frac{L}{22n}}}
\end{equation}  
and by choosing the appropriate $n$, the Bell pair generation scales at long distances.
In the quantum Internet, the link Bell pairs created at each link in the path are ''connected'' to generate an end-to-end Bell pair, and quantum teleportation is performed using the connected Bell pair to realize arbitrary quantum data transmission between end nodes.

\subsection{Entanglement purification}
\label{subsec:ep}
\begin{figure}[t].
\begin{center}
    \includegraphics[width=8cm]{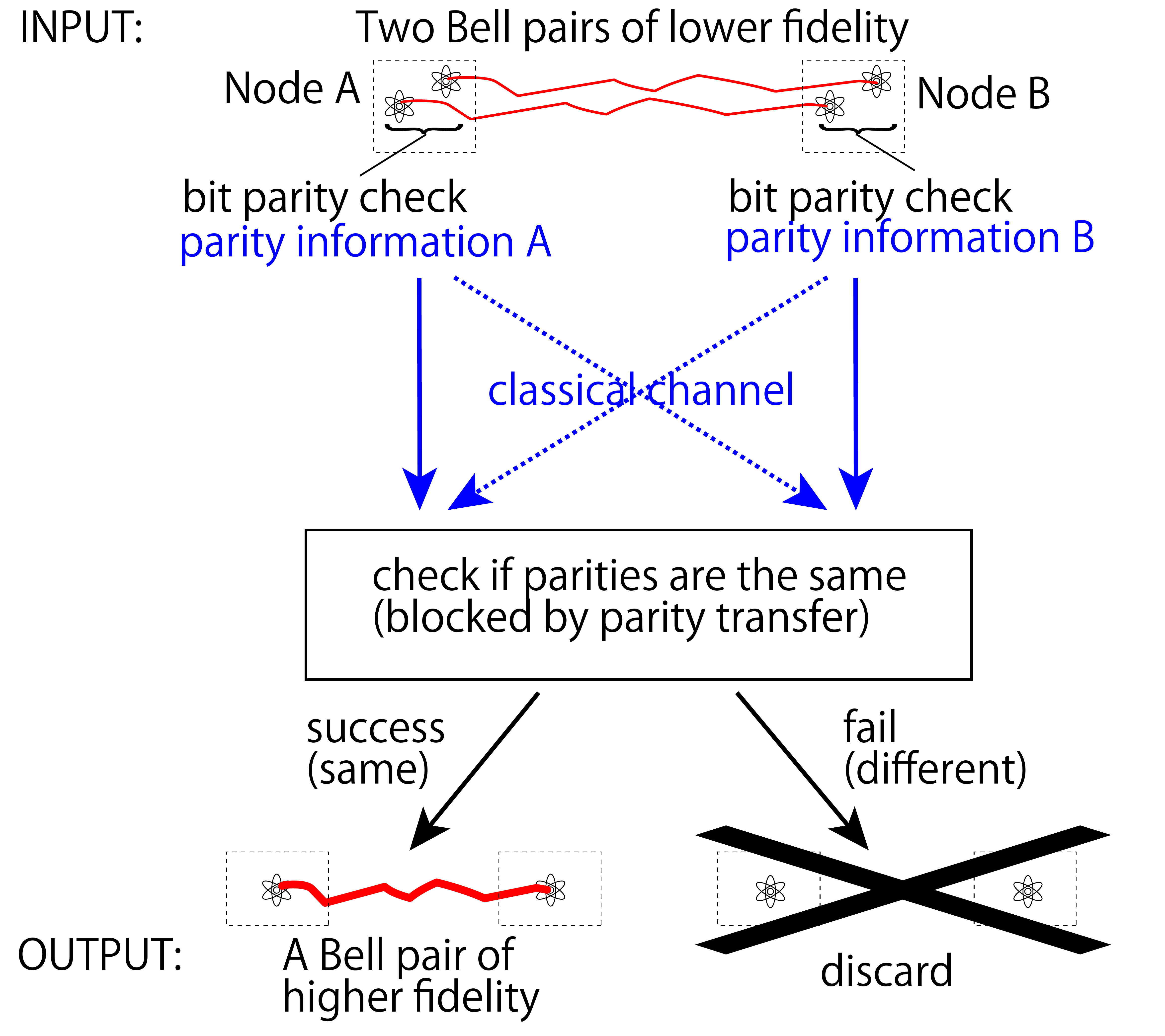}
    \caption{Figure of an Entanglement Purification. 
    Success and failure are probabilistic.
    The higher the input fidelities are, the higher the success probability is.}
    \label{fig:puri}
\end{center}
\end{figure}

An important feature of quantum Internet is error management.
An algorithm to improve the fidelity of Bell pairs by error detection is called entanglement purification.
An image of the most realistic of the several purification algorithms is shown in Fig. \ref{fig:puri}~\cite{Briegel1998}.
Since this error detection also consumes a bell pair, purification is an algorithm that consumes one set of Bell pairs to improve the fidelity of another set.
In addition, the fidelity is improved by discarding the error-detected bell pairs.
Therefore, entanglement purification is a stochastic algorithm that generates a Bell pair with high fidelity only in case of success.

The error detection mechanism is parity detection.
In the absence of errors, a half Bell pair in Node A is 0 (1) if the corresponding half Bell pair in Node B is 0 (1).
Therefore, when we have two sets of Bell pairs shared between Node A and B, the bit parity of the two qubits held by Node A and the bit parity of the two qubits held by Node B are identical, although we don't know the specific oddity.
The quantum state in the error-free case is
\begin{align}\label{equ:puri_no_error}
  \vert \Phi^+_{01} \rangle \otimes \vert \Phi^+_{23} \rangle =
  \frac{1}{2}(&
    \vert 0_00_1 0_20_3 \rangle +
    \vert 0_00_1 1_21_3 \rangle + \nonumber \\
    &\vert 1_01_1 0_20_3 \rangle +
    \vert 1_01_1 1_21_3 \rangle
  ).
\end{align}

Eq. ~\ref{equ:puri_no_error}, the qubit $0$ and qubit $2$ are in Node A, and the qubit $1$ and qubit $3$ are in Node B.
The parity of qubit $0$ and qubit $2$ and the parity of qubit $1$ and qubit $3$ are identical.
As a case where there is an error, for example, consider the case where the qubit $1$ is bit-flipped.
\begin{align}\label{equ:puri_error}
  \vert \Psi^+_{01} \rangle \otimes \vert \Phi^+_{23} \rangle =
  \frac{1}{2}(&
    \vert 0_01_1 0_20_3 \rangle +
    \vert 0_01_1 1_21_3 \rangle + \nonumber \\
    &\vert 1_00_1 0_20_3 \rangle +
    \vert 1_00_1 1_21_3 \rangle
  )
\end{align}
In this case, the parity of qubit $0$ and qubit $2$ does not match that of qubit $1$ and qubit $3$.
This parity detection is used to detect errors, but it is not known whether the error was in the Bell pair consumed for parity detection or the remaining Bell pair.
Therefore, when an error is detected, it is necessary to discard the residual Bell pair.
Note that error cannot be detected if there is a bit-flip error in both sets of Bell pairs.

As described above, the algorithm must inform each other of the observed parity to determine success or failure.
Since one bit of information is sent to each other for each purification, the consumption of classical communication resources is not significant. Still, there is a problem that quantum memory is occupied while waiting for the measurement results to arrive from the other node, regardless of success or failure.  
Since this is a consumption of communication resources, it becomes a scalability problem.
We discuss this issue in detail in Sec.~\ref{sec:problem}.

To detect phase-flip errors, the Hadamard gate is used to exchange the bit, and phase components of the qubit, and purification is performed again.
Therefore, the fidelity of the Bell pair is greatly improved when the purification is performed twice. (The fidelity itself is a single indicator that incorporates both bit-flip and phase-flip errors.)

\section{Problem}
\label{sec:problem}
\begin{figure*}[t].
  \begin{center}
    \includegraphics[width=17cm]{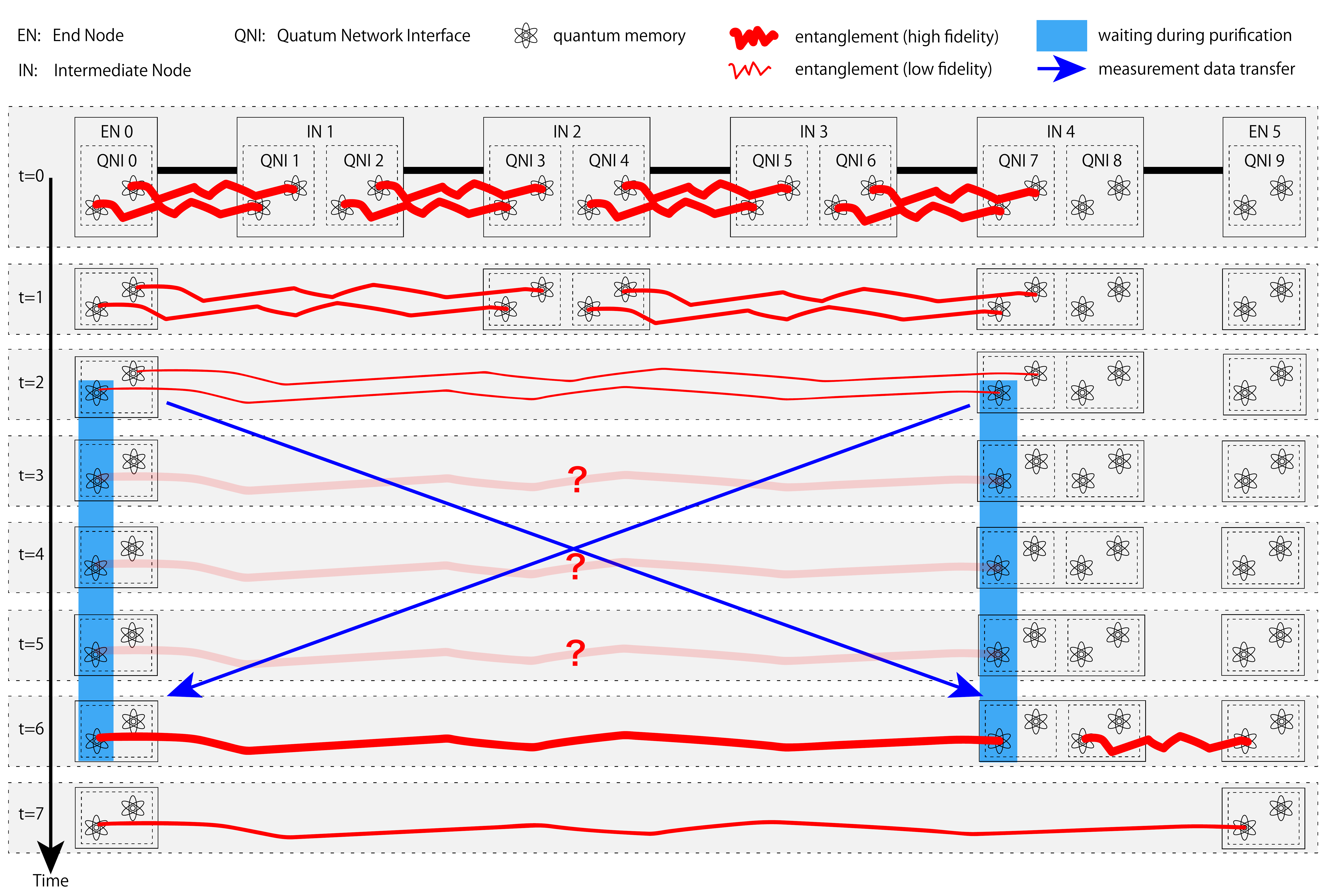}
    \caption{A picture that depicts the workload problem of entanglement purification.
    Each end node (EN) has a quantum network interface (QNI), and each intermediate node (IN) has two QNIs.
    Intermediate nodes can operate entanglement swapping between different QNIs.
    Every node can operate entanglement purification between their qubits.
    For visibility, two qubits are shown in each QNI. QNI may have more qubits.
    Entanglement purification occpies quantum memory during classical communication (blue box).
    Blocking qubits in intermediate nodes is a scalability issue.
    }
    \label{fig:problem}
  \end{center}
\end{figure*}

The problem to be solved in this study is displayed in Fig. ~\ref{fig:problem}.
Between two end nodes (EN0 and EN5), there are four intermediate nodes (IN1, IN2, IN3, and IN4). 
These are connected by optical fiber, etc., and a link Bell pair can be generated by entangling the quantum memory between adjacent nodes by sending photons ($t=0$).
If the fidelity of the raw link Bell pair is low, link-level entanglement purification can be performed to generate link Bell pairs with adequate high fidelity.
The Bell pairs generated between neighboring nodes can be "connected" by entanglement swapping ($t=1$).
However, since the fidelity of both Bell pairs before the connection is inherited, the fidelity is reduced.
Entanglement swapping can be repeated to achieve a Bell pair between more distant nodes, but the fidelity drops further ($t=2$).
To increase the fidelity, entanglement purification is performed (purification at the internetworking level).
When entanglement purification is performed, the nodes have to transmit the observed parity to each other over the classical channel (probably the Internet) to compare the parities at both nodes (EN0 and IN4), resulting in a latency ($t=2 \sim 6$).
If the collation is successful, a Bell pair with good fidelity is output ($t=6$).
This waiting time is a load on the network, and the farther the distance, the greater the load because the transmission time of the observed parities increases.
Moreover, during this waiting time, the quantum memory is also subject to the decoherence effect, which increases the need for further purification.

Note that if the number of hops $n$ is not $2^m$ of somewhat $m$, the connection by entanglement swapping has a "remainder" like the Bell pair (QNI8-QNI9).
If the quantum communication request requires many Bell pairs and Bell pairs are being created continuously at a sufficiently high rate, some Bell pairs (QNI0-QNI7) and some Bell pairs (QNI8-QNI9) may be generated at the same time.
Therefore, we can quickly move on to the next entanglement swapping ($t=7$).
If the fidelity of the Bell pair (QNI0-QNI9) is insufficient, entanglement purification can also be performed by preparing two sets of end-to-end Bell pairs (end-to-end level entanglement purification).

\section{End-to-End Principle for Entanglement Purification}
\label{sec:e2eep}

\begin{figure*}[t]
    \begin{center}
      \includegraphics[width=17cm]{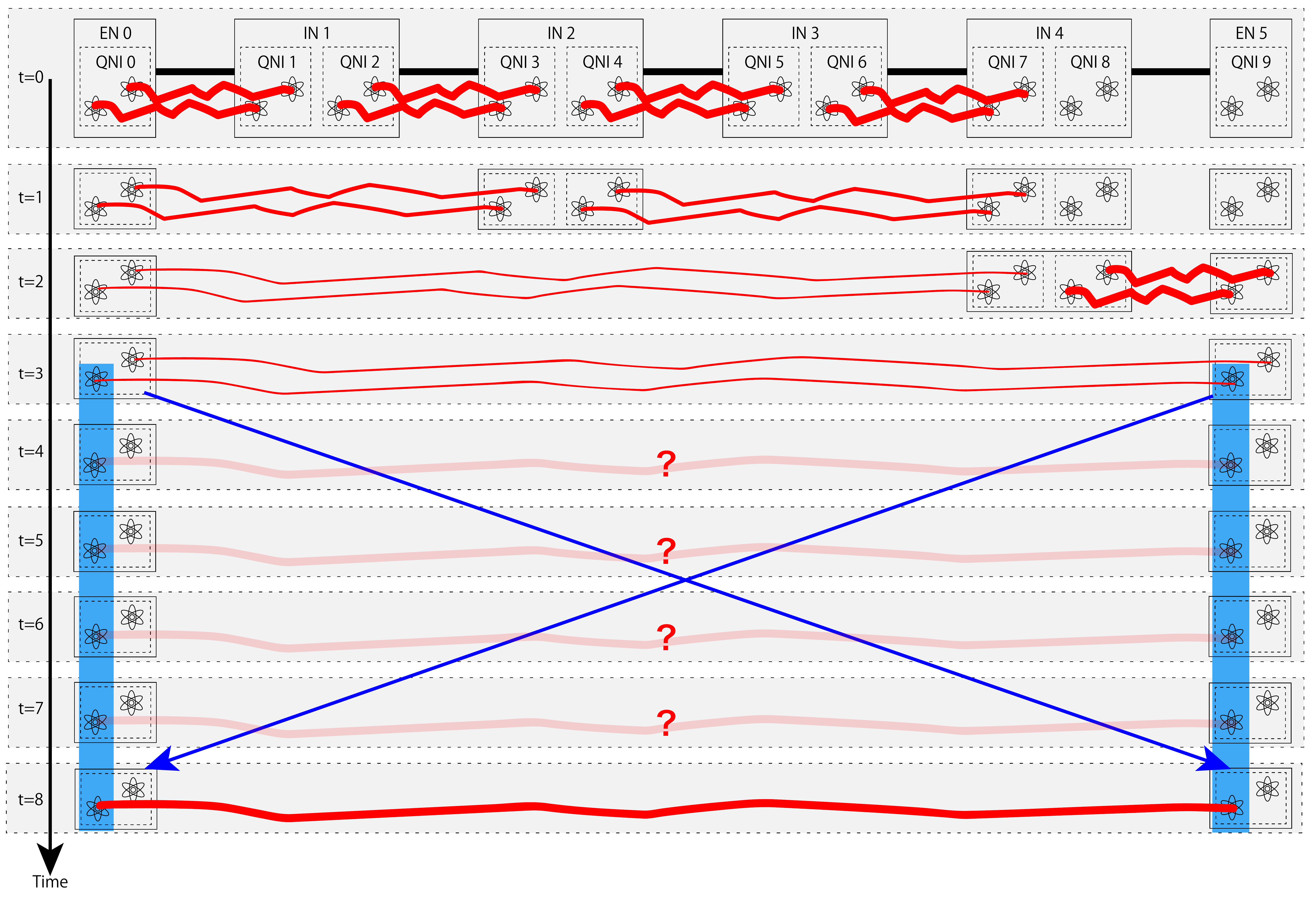}
      \caption{Entanglement Purification Architecture which does not let intermediate nodes execute non-local purification.}
      \label{fig:solution}
    \end{center}
  \end{figure*}

To reduce the network load in the communication, we propose to perform the purification at the local link level or the end-to-end level.
The overview is illustrated in Fig. ~\ref{fig:solution}.
In Fig. ~\ref{fig:problem}, non-local purification was also performed by intermediate nodes.
In our proposal, as illustrated in Fig. ~\ref{fig:solution}, non-local purification is performed only at the end node.
This scheme is expected to reduce the amount of time of quantum memory that the intermediate node has to devote to, thus reducing the load on the network side and improving the scalability of the quantum Internet.
On the other hand, of course, end-to-end Bell pairs are generated by consuming network resources, so to perform end-to-end purification, the load on the network is devoted to creating the two sets of Bell pairs as input.
Therefore, the trade-off of reducing and increasing this load determines the effect in terms of resource efficiency. 
In this work, we examine this resource efficiency by simulations.

\section{Simulation}
In this work, we used simulations to calculate the total qubit occupation time in the network (intermediate node) and end node that occurs when generating one end-to-end quantum entanglement with the required fidelity in the fashion in Fig. ~\ref{fig:problem} and Fig. ~\ref{fig:solution}, respectively. 
Because the occupation time depends on fidelity due to purification, we traced both fidelity and total occupation time along with simulating network operations.

\subsection{Model}
\begin{figure}
  \begin{center}
    \includegraphics[width=8cm]{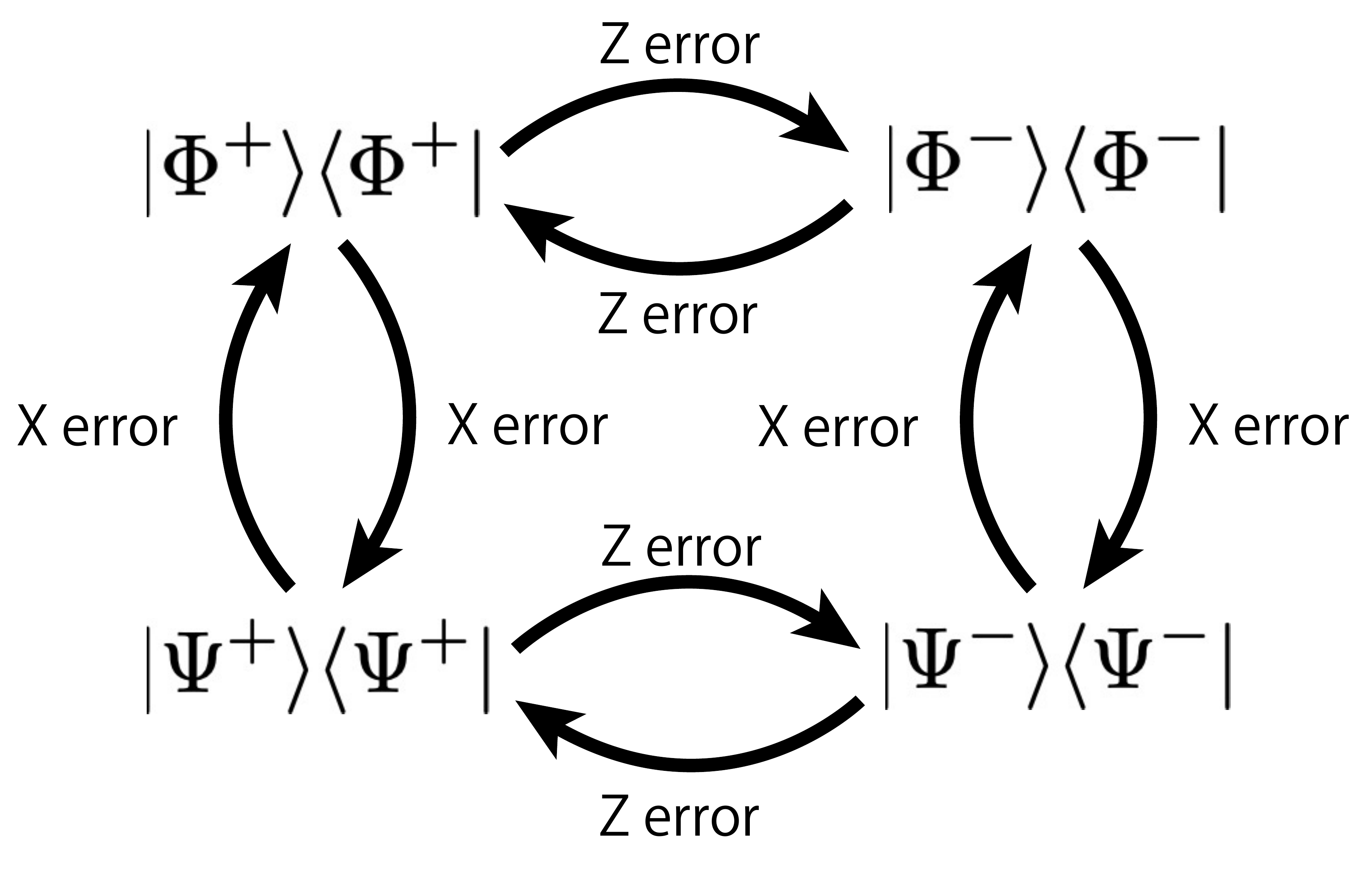}
    \caption{State transtion among non-errored and errord states by errors.}
    \label{fig:state_transition}
  \end{center}
\end{figure}

In this study, we treat Eq.~\ref{equ:state1} as a model of a quantum state.
At each time step, the state changes according to the error probability $p$. The same state is maintained (identity gate) with probability $1-p$, and $\frac{p}{3}$ generates $X$, $Y$ and $Z$ errors (unintentional $X$, $Y$ and $Z$ gates, respectively). The state transition diagram is shown in Fig. \ref{fig:state_transition}.
When the state at time $t$ is described as
\begin{equation}
  \rho_t = A_t \vert \Phi^+ \rangle \langle \Phi^+ \vert +
         B_t \vert \Psi^+ \rangle \langle \Psi^+ \vert + 
         C_t \vert \Psi^- \rangle \langle \Psi^- \vert + 
         D_t \vert \Phi^- \rangle \langle \Phi^- \vert,
  \label{equ:state1}
\end{equation}
the state at time $t+1$ is described as 
\begin{align}
  \rho_{t+1} = 
  &\{(1-p)A_t + \frac{p}{3}(B_t + C_t + D_t)\} \vert \Phi^+ \rangle \langle \Phi^+ \vert + \nonumber \\
  &\{(1-p)B_t + \frac{p}{3}(A_t + C_t + D_t)\} \vert \Psi^+ \rangle \langle \Psi^+ \vert + \nonumber \\
  &\{(1-p)C_t + \frac{p}{3}(A_t + B_t + D_t)\} \vert \Psi^- \rangle \langle \Psi^- \vert + \nonumber \\
  &\{(1-p)D_t + \frac{p}{3}(A_t + B_t + C_t)\} \vert \Phi^- \rangle \langle \Phi^- \vert .
  \label{equ:states_after}
\end{align}

\subsubsection{Errors on Entanglement Swapping}
The output Bell pairs' errors depend on the errors that the two sets of Bell pairs input to entanglement swapping have.
Tab. ~\ref{tab:es_output} lists the relationships of input and output.
\begin{table}
  \begin{center}
    \begin{tabular}{c|c|c}
      input 1 & input 2 & output\\
      \hline
      $\vert \Phi^+ \rangle$ & $\vert \Phi^+ \rangle$ & $\vert \Phi^+ \rangle$ \\
      $\vert \Phi^+ \rangle$ & $\vert \Psi^+ \rangle$ & $\vert \Psi^+ \rangle$ \\
      $\vert \Phi^+ \rangle$ & $\vert \Phi^- \rangle$ & $\vert \Phi^- \rangle$  \\
      $\vert \Phi^+ \rangle$ & $\vert \Psi^- \rangle$ & $\vert \Psi^- \rangle$  \\

      $\vert \Psi^+ \rangle$ & $\vert \Phi^+ \rangle$ & $\vert \Psi^+ \rangle$ \\
      $\vert \Psi^+ \rangle$ & $\vert \Psi^+ \rangle$ & $\vert \Phi^+ \rangle$ \\
      $\vert \Psi^+ \rangle$ & $\vert \Phi^- \rangle$ & $\vert \Psi^- \rangle$  \\
      $\vert \Psi^+ \rangle$ & $\vert \Psi^- \rangle$ & $\vert \Phi^- \rangle$  \\

      $\vert \Phi^- \rangle$ & $\vert \Phi^+ \rangle$ & $\vert \Phi^- \rangle$  \\
      $\vert \Phi^- \rangle$ & $\vert \Psi^+ \rangle$ & $\vert \Psi^- \rangle$  \\
      $\vert \Phi^- \rangle$ & $\vert \Phi^- \rangle$ & $\vert \Phi^+ \rangle$ \\
      $\vert \Phi^- \rangle$ & $\vert \Psi^- \rangle$ & $\vert \Psi^+ \rangle$ \\

      $\vert \Psi^- \rangle$ & $\vert \Phi^+ \rangle$ & $\vert \Psi^- \rangle$  \\
      $\vert \Psi^- \rangle$ & $\vert \Psi^+ \rangle$ & $\vert \Phi^- \rangle$  \\
      $\vert \Psi^- \rangle$ & $\vert \Phi^- \rangle$ & $\vert \Psi^+ \rangle$ \\
      $\vert \Psi^- \rangle$ & $\vert \Psi^- \rangle$ & $\vert \Phi^+ \rangle$
    \end{tabular}
    \caption{Input states and their output of entanglement swapping.}
    \label{tab:es_output}
  \end{center}
\end{table}
In addition to this state change, the quantum operation of entanglement swapping itself is imperfect hence deteriorates the fidelity.
Therefore, in simulating the error in entanglement swapping, after considering the state change in Tab. ~\ref{tab:es_output}, a probabilistic error process described in Eq. ~\ref{equ:states_after} is applied depending on the operation error probability $p_{op}$ of Node B in Fig. ~\ref{fig:es}, and ones depending on the memory error probability $P_{mem}$ that Nodes A and C have are applied.
These operations are performed in one unit of time, which occupies four quantum memories.
The feedback of the parity information of entanglement swapping can be reflected after the latter operation is performed if the subsequent operation is entanglement swapping or entanglement purification.
Therefore, the transmission latency of the parity information of entanglement swapping is not considered in this study.

\subsubsection{Errors on Purification}
As the same, the output Bell pairs' errors depend on the errors that the two sets of Bell pairs input to entanglement purification have.
Tab. ~\ref{tab:ep_output} lists the relationships of input and output.
\begin{table}
  \begin{center}
    \begin{tabular}{c|c|c}
      input 1 & input 2 & output\\
      \hline
      $\vert \Phi^+ \rangle$ & $\vert \Phi^+ \rangle$ & $\vert \Phi^+ \rangle$ \\
      $\vert \Phi^+ \rangle$ & $\vert \Psi^+ \rangle$ & discard\\
      $\vert \Phi^+ \rangle$ & $\vert \Phi^- \rangle$ & discard \\
      $\vert \Phi^+ \rangle$ & $\vert \Psi^- \rangle$ & $\vert \Psi^- \rangle$ \\

      $\vert \Psi^+ \rangle$ & $\vert \Phi^+ \rangle$ & discard \\
      $\vert \Psi^+ \rangle$ & $\vert \Psi^+ \rangle$ & $\vert \Psi^+ \rangle$ \\
      $\vert \Psi^+ \rangle$ & $\vert \Phi^- \rangle$ & $\vert \Phi^- \rangle$ \\
      $\vert \Psi^+ \rangle$ & $\vert \Psi^- \rangle$ & discard \\

      $\vert \Phi^- \rangle$ & $\vert \Phi^+ \rangle$ & discard \\
      $\vert \Phi^- \rangle$ & $\vert \Psi^+ \rangle$ & $\vert \Phi^- \rangle$ \\
      $\vert \Phi^- \rangle$ & $\vert \Phi^- \rangle$ & $\vert \Psi^+ \rangle$ \\
      $\vert \Phi^- \rangle$ & $\vert \Psi^- \rangle$ & discard \\

      $\vert \Psi^- \rangle$ & $\vert \Phi^+ \rangle$ & $\vert \Psi^- \rangle$ \\
      $\vert \Psi^- \rangle$ & $\vert \Psi^+ \rangle$ & discard \\
      $\vert \Psi^- \rangle$ & $\vert \Phi^- \rangle$ & discard \\
      $\vert \Psi^- \rangle$ & $\vert \Psi^- \rangle$ & $\vert \Phi^+ \rangle$ 
    \end{tabular}
    \caption{Input states and their output of entanglement purification.}
    \label{tab:ep_output}
  \end{center}
\end{table}
Similar to entanglement swapping, the quantum operation of entanglement purification itself worsens fidelity.
Therefore, in the simulation of errors in purification, after considering the state change in Tab.~\ref{tab:ep_output}, a probabilistic error process described in Eq. ~\ref{equ:states_after} is applied depending on the operation error probability $p_{op}$ of Node A and B in Fig. ~\ref{fig:ep}.
These operations are performed in one unit of time, occupying four quantum memories.
Unlike Entanglement swapping, the subsequent operations cannot be executed first because the transmission of the parity information of the purification is related to the success or failure of the purification.
For the sake, we apply an Eq. ~\ref{equ:states_after} operation depending on the memory error probability $p_{mem}$ of Node A and B for the transmission time proportional to the number of hops $d$ between A and B.
We also accounted for an additional unit time of $d$ times occupying two quantum memories as resource consumption for this transmission latency.

\subsubsection{Success rate of entanglement purification}
As mentioned earlier, entanglement purification may discard output Bell pairs.
From a statistical point of view, the success probability is described depending on the original error state.
In this study, we compared Eq.~\ref{equ:state1} and Tab.~\ref{tab:ep_output}, and when there is two sets of input Bell pair $0$ and $1$, we set $(A_0+D_0)*(A_1+D_1)+(B_0+C_0)*(B_1+C_1)$ as the success probability. 
The number of resources consumed to generate the output Bell pair was calculated by multiplying the number of resources consumed to generate the input Bell pair by the inverse of this success probability.

\subsection{Assumptions and Variables}
In the simulations of this study, we make the following assumptions and variables.
\begin{itemize}
  \setlength{\itemsep}{0pt} 
	\setlength{\parskip}{2pt} 
  \item {\em raw Bell pair generation success rate:} The probability of succeeding in generating a link-level Bell pair, taking into account Photon loss. We set it to 0.01. This value is better than the current experimental results but is considered within the range of feasibility in the future. This variable does not affect the comparison of the purification methods in this work.
  \item {\em unit time:} 0.0001 seconds. Equivalent to the time it takes for light to travel approximately 20 km in an optical fiber.
  \item {\em raw Bell pair fidelity:} We set this value to 0.8, after experimental results 0.85 in a laboratory~\cite{Nolleke2013}.
  \item {\em link-level fidelity threshold (L2 fidelity threshold. "L2" is temporarily named after the classical OSI model for the space in the legends of the graph.):} If the fidelity of the link-level Bell pair is below this threshold, entanglement purification is recursively performed until the fidelity gets above this threshold. We tested three patterns: 0.90, 0.99, and 0.999.
  \item {\em internetworking-level fidelity threshold (L3 fidelity threshold.):} 
  If the fidelity of the internetworking-level (remote intermediate nodes level) Bell pair is below this threshold, perform purification (recursively). We tested this for three patterns: 0.90, 0.99, and 0.999. For end-to-end purification, this threshold does not work; therefore, "None" is shown in Fig.~\ref{fig:graph}.
  \item {\em end-to-end-level fidelity threshold (L4 fidelity threshold):} 
  If the fidelity of the end-to-end Bell pair is below this threshold, perform purification (recursively). This value is also the fidelity of the Bell pair passed to the application side by the end node. In this study, we set this value to 0.99.
  \item {\em memory error rate at intermediate node:} 
  The probability of one of the X/Z/Y errors occurring in a qubit in intermediate nodes. We set this value 0.0001 (since the unit time is 0.0001 seconds, the fidelity falls below 0.5 in 0.8240 seconds.) or 0.00001, depending on the experiment.
  \item {\em operation error rate at intermediate node:} 
  The probability of error in operation in a single intermediate node in entanglement swapping or entanglement purification. 
  When we run entanglement swapping, we apply this error probability once for the center node.
  When we run entanglement purification, we apply this error probability separately for the left and right nodes.
  We use the same value with the memory error rate at the intermediate node in this study.
  \item {\em memory error rate at end node:} 
  The same as the memory error rate but at end nodes.
  We set this value 0.0001, 0.00001, or 0, depending on the experiment.
  We try multiple End Node capabilities and compare the results.
  \item {\em operation error rate at end node:} Same as above.
\end{itemize}

\subsection{Result}
\begin{figure*}
  \begin{center}
      (a)\includegraphics[width=8.3cm]{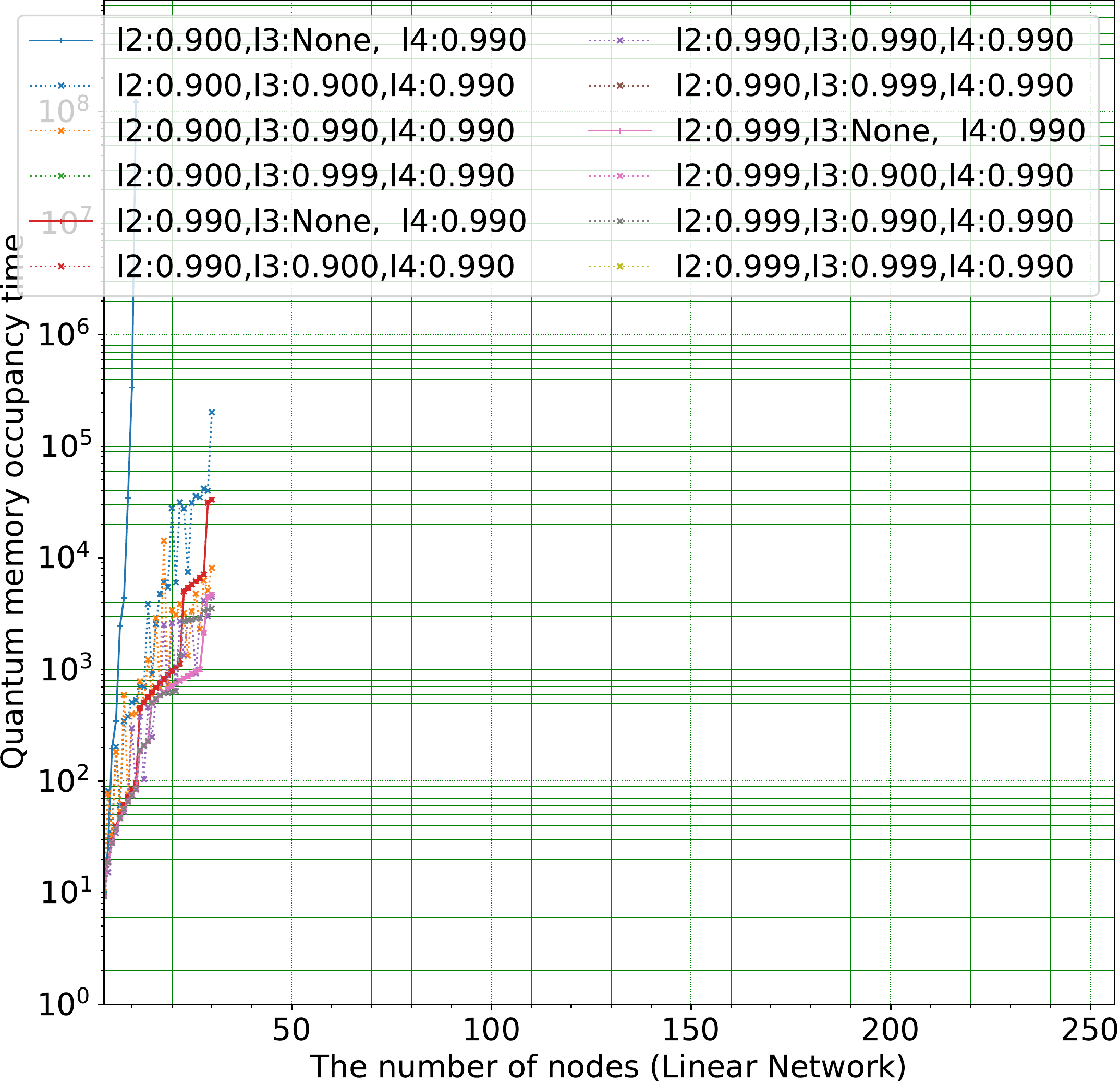}
      (b)\includegraphics[width=8.3cm]{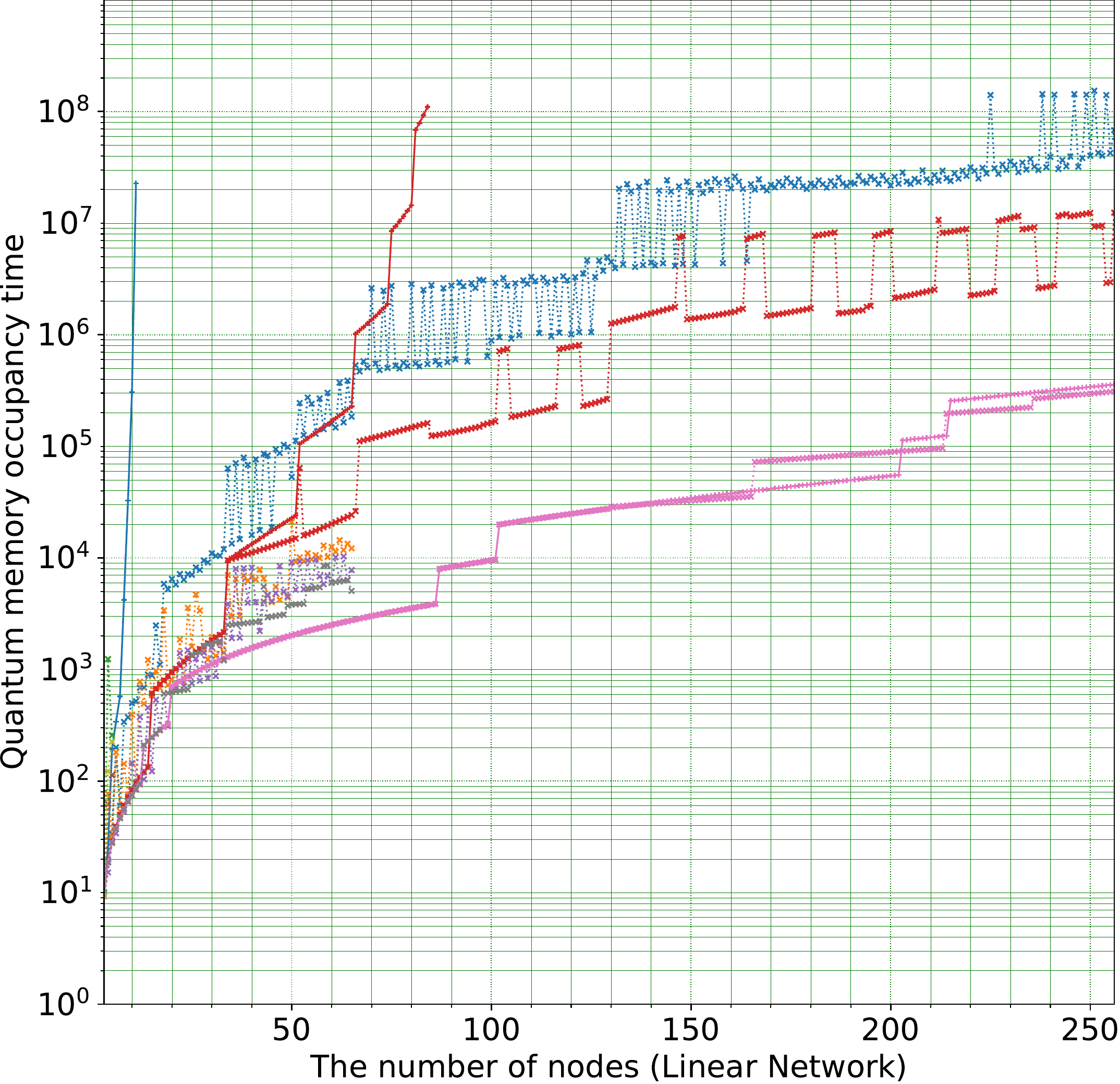}\\
      \vspace{3mm}
      (c)\includegraphics[width=8.3cm]{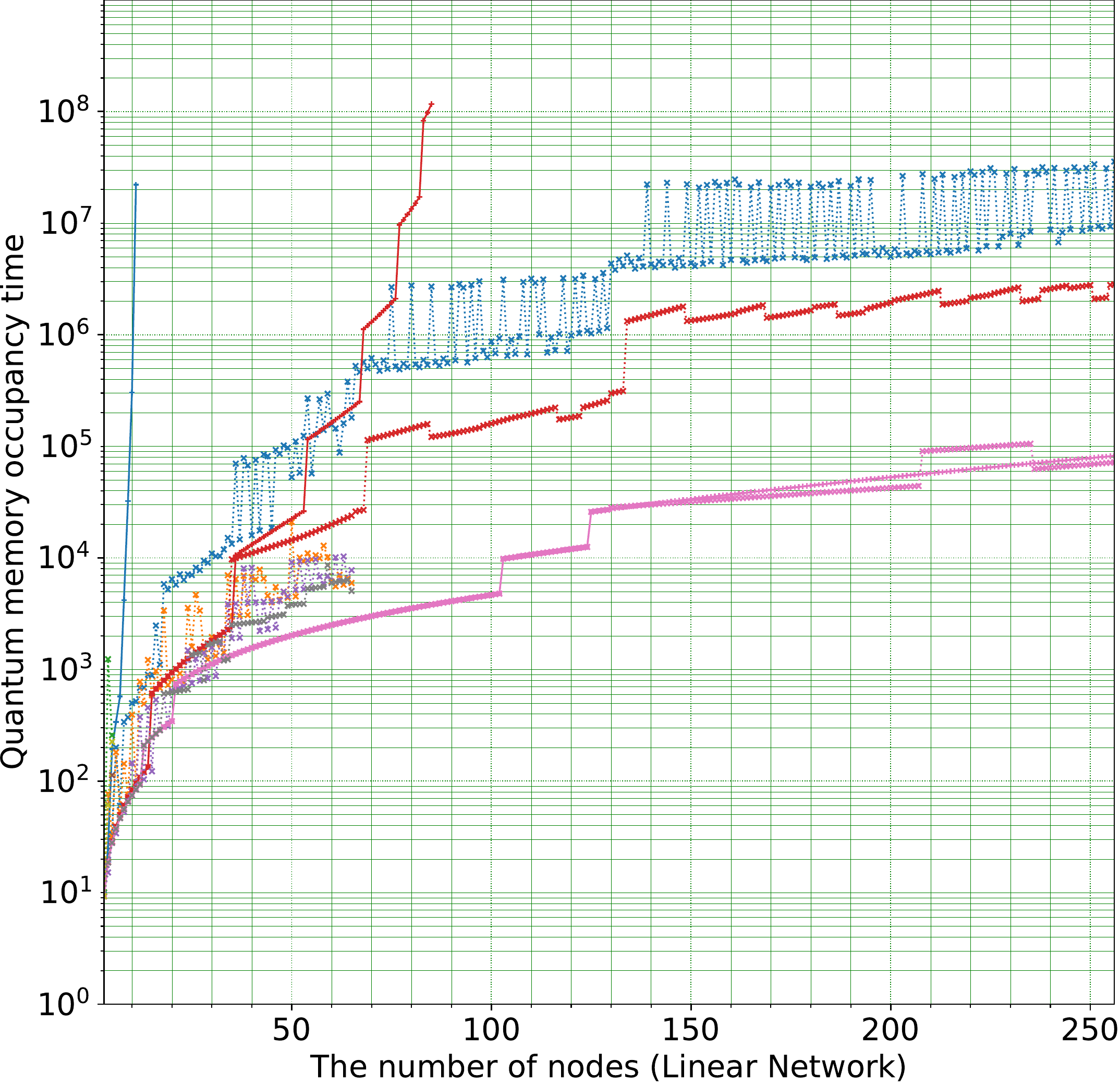}
      (d)\includegraphics[width=8.3cm]{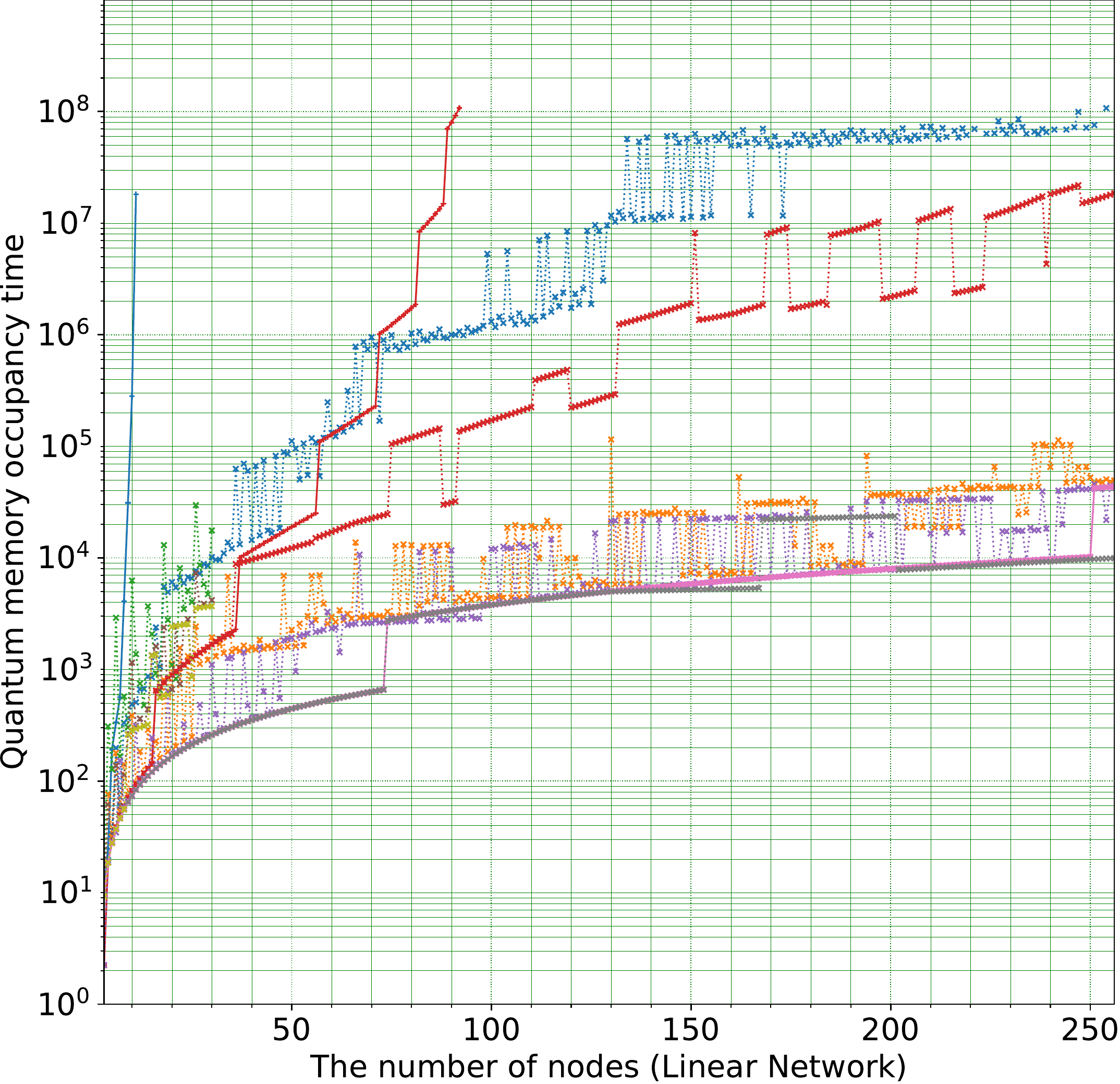}
  \caption{Graphs which depicts the relationship between the number of hops and the total load on the intermediate nodes. The legends are in common.
  (a) The error probabilities of the end nodes are $0.0001$, the same as that of the intermediate node.
  (b) The error probabilities of the end nodes are $0.00001$, lower than that of the intermediate node, suggesting that a physical system such as a Trapped-ion is used for the qubit of the end node. Trapped-ions have a low error rate but take a long time for quantum gate operations, so end nodes may be more suitable than intermediate nodes that need to handle many requests.
  (c) The error probabilities of the end node are ultimately lower, $0$; e.g., a fully operational fault-tolerant quantum computer may be assumed. Intermediate nodes have error rate of $0.0001$, meaning that they don't have fully fault-tolerant system which requires heavy quantum resources.
  (d) Error probabilities for both intermediate node and end node is $0.00001$. Almost no error occurs.}
  \label{fig:graph}
  \end{center}
\end{figure*}

Fig. ~\ref{fig:graph}(a), ~\ref{fig:graph}(b), ~\ref{fig:graph}(c), and ~\ref{fig:graph}(d) illustrate the simulation results.
In these figures, to consider the load on the network, we simulate the total time of qubits dedicated by the intermediate node to generate one end-to-end Bell pair with a specific fidelity (0.990) relative to the number of nodes on the path between the end nodes~\footnote{Graphs related to end nodes and in total are provided in appendix. Actually their shape of graphs are totally same as that of intermediate nodes.}.
The qubit occupation time of the end node and the total occupation time of the two nodes are shown in Sec. ~\ref{appendix:mr} of the appendix.
Overall, it can be seen that setting a high fidelity threshold (0.999) at the L2 level first and performing purification to increase the fidelity of the link Bell pair contributes to the efficiency.

In Figure ~\ref{fig:graph}(a), the graph is broken at 30 nodes in every settings. This result means that although we can generate an end-to-end Bell pair, we cannot generate an end-to-end Bell pair with the target fidelity of 0.99 because the error probability at the end node is too high, and the fidelity deteriorates during the classical communication time of 29 hops to check the purification result~\cite{Hartmann2007}.

In Fig. ~\ref{fig:graph}(b), the two settings (l2:0.999, l3:None, l4:0.990) and (l2:0.999, l3:0.900, l4:0.990) are more efficient. 
The difference between the two is that the former performs purification only at local and end-to-end, while the latter also performs internetworking-level purification with a threshold of 0.9. 
The graphs are consistent up to 129 nodes because internetworking-level purification is not performed because the threshold of 0.900 is too low, and thus it is effectively end-to-end purification. 
In addition, since the error probability at the end node is low, it is presumed that it is more efficient to purify after connecting the Bell pair to the end node than to purify at the intermediate node.
(l2:0.999, l3:0.990, l4:0.990) is broken at 65 nodes, which is much better than the 30 nodes that resulted from a similar setting in Fig. ~\ref{fig:graph}(a). 
This result can be attributed to the fact that the fidelity decoherence at the end node, which holds the half Bell pair for a long time as shown by QNI0 in Fig. ~\ref{fig:problem}, is slower in this experiment.

In Fig. ~\ref{fig:graph}(c), we further reduce the error probability of the end nodes to 0.
Comparing (l2:0.999, l3:None, l4:0.990) and (l2:0.999, l3:0.900, l4:0.990) in Fig. ~\ref{fig:graph}(c) with (l2:0.999, l3:0.900, l4:0.990) in Fig. ~\ref{fig:graph}(b), the load of intermediate nodes gets lower although we reduce only the error probability at the end nodes.
The reason should be that the improvement of the error probability in the end nodes improves the efficiency of purification, reducing the load on the intermediate node that is purifying together with the end node.
In reality, a fault-tolerant quantum computer is necessary to achieve the error probability to be 0. However, the qubits used for computation in a fault-tolerant quantum computer are created by quantum error correction codes consisting of thousands of qubits and requiring tens of error-correcting cycles to initializing a state. In this study, to consider the extreme setting, we also simulated this setting where the error probability is zero without considering the error correction code.

In Fig. ~\ref{fig:graph}(d), the error probability is set to 0.00001 to avoid the overhead in communication distance caused by the purification threshold.
Still, (l2:0.999, l3:0.999, l4:0.990) breaks in 30 nodes due to the high internetworking-level threshold.
(l2:0.999, l3:None, l4:0.990) and (l2:0.999, l3:0.900, l4:0.990), as well as (l2:0.999, l3:0.990, l4:0.990), have similar graphs.
As the number of nodes increases, they are expected to show the same relationship as the number of nodes 130 onwards  (l2:0.999, l3:None, l4:0.990) (l2:0.999, l3:0.900, l4:0. 990) in Fig. ~\ref{fig:graph}(b).
The graphs of (l2:0.900, l3:0.990, l4:0.990) and (l2:0.990, l3:0.990, l4:0.990) have also similar values. These graphs show that purification is performed by L3's threshold at a distance of a few hops, increasing somewhat local purification. This results in that purification are not performed at long distances.

\begin{figure}[t]
  \begin{center}
    \includegraphics[width=8cm]{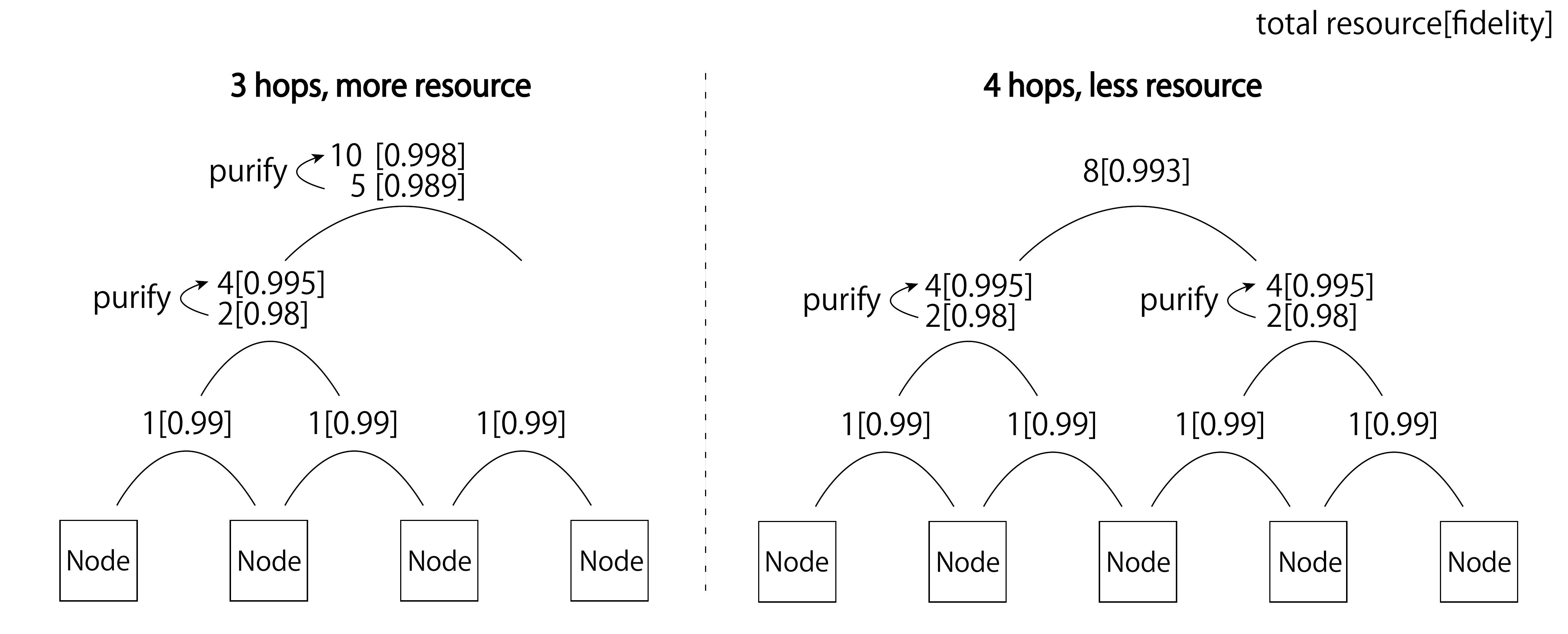}
    \caption{A simplified example which explains cases in which a connection of less hops needs more resource. 
    The thresholds to execute purification is 0.99. The numbers are example for simplicity.}
    \label{fig:zigzag}
  \end{center}
\end{figure}

Some of the non-end-to-end data is zigzagged.
This is because purification between more distant nodes consumes more resources and also because there is a threshold for the execution of purification.
A simple explanation is shown in Fig. ~\ref{fig:zigzag}.
By increasing the number of hops by 1 hop, we are able to purify nodes within a distance of 2 hops and generate end-to-end Bell pairs with fewer resources.
Such a sudden high resource demand may be fixed by optimizing the scheduling of purification carefully than just obeying the threshold.

\section{Discussion}
\label{sec:discussion}
\begin{figure}[t]
    \includegraphics[width=8cm]{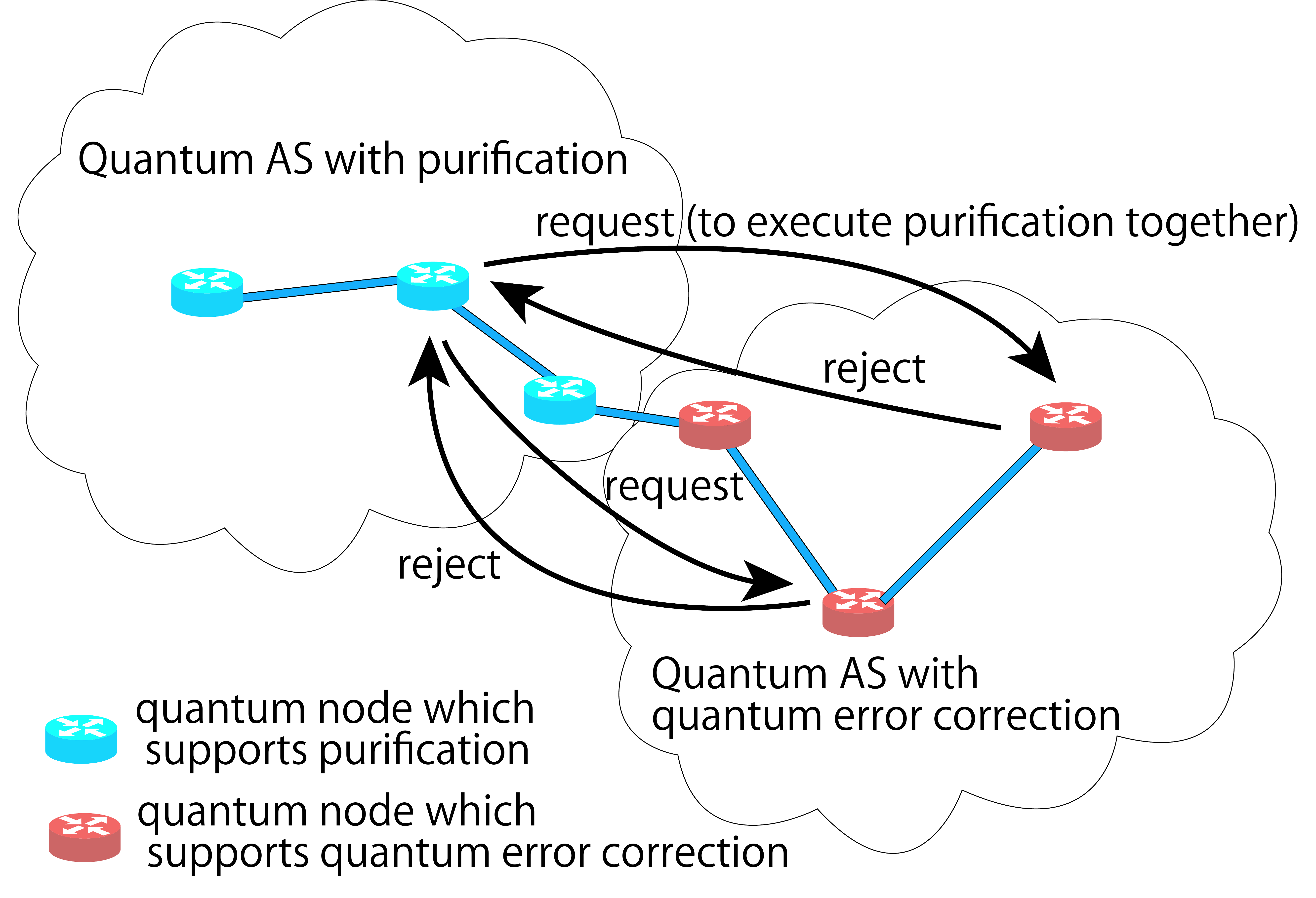}
    \caption{Possible problem which occurs when neighboring quantum Autonomous Systems (ASs) employ different error management schemes.}
    \label{fig:qas}
\end{figure}

Non-local processing may cause problems in the interoperability of the quantum Internet.
Fig. ~\ref{fig:qas} illustrates a possible problem when two neighboring Quantum Autonomous Systems (AS) use different error management schemes~\cite{Muralidharan2016}.
A quantum node in a quantum AS that employs entanglement purification may request to perform purification with a quantum node in another quantum AS that employs Quantum Error Correction (QEC) for the error management scheme, but the request gets rejected due to the mismatch of error management scheme.
Technically it is possible to interoperate different error management schemes in a single node~\ref{Nagayama2016}.
However, it may cause interoperability or scalability problems if every node has to support many schemes or if every node has to know other nodes' supporting error management schemes.
This problem can also occur between non-adjacent quantum ASes.
Even on a network, if the error management scheme is locally closed or end-to-end error management principle is employed, this problem does not occur because it can be solved by configuration between two parties, locally or end-to-end.
Therefore, purification using the end-to-end principle will contribute to ensuring the autonomous decentralized cooperation, interoperability, and scalability of the quantum Internet.

Our study's unit time of 0.0001 seconds takes a photon to travel 20 km in an optical fiber. 
Therefore, our 1 hop is 20km, and 256 nodes correspond to 5100km.
Since the earth's circumference is about 40,000 km, this is equivalent to 1/8 of the earth.
Our simulation shows that the distance over which fidelity can be improved by purification depends on the error probability of the qubit in the end node.
It may be challenging to use purification to generate a Bell pair with the other side of the world.
In such a case, a Bell pair with a fidelity sufficiently higher than the desired fidelity can be generated at intermediate nodes within the range not exceeding the distance limit and connected by entanglement swapping, without waiting classical communication "blindly", termed by ~\cite{Hartmann2007}.
In such a case, blocking qubits in the intermediate node, a vital communication resource, recurs.
To solve this problem, we can attach a "side box" to intermediate nodes to hold half Bell pairs for a long time, generate the bell pair in the end node and the side box, and perform purification there.

\section{Conclusion}
\label{sec:conc}
It is a well-known rule of computer networking that lightening the load on intermediate nodes in individual communication processing improves the communication capacity of the network.
In other words, by executing tasks that can be performed by end nodes as much as possible, intermediate nodes such as routers can perform more communication processing, thereby securing the communication capacity of the network.
In this study, we proposed an architecture in which the intermediate nodes do not perform high-load purifications.
As a result of the simulation, the resources on the network side can be used more efficiently when there is an end node specializing in error management.
In this sense, to equip a module dedicated for entanglement purification to end nodes is helpful, and qubits for computation or sensing do not need to have such specialized capability.
In addition, we confirmed that even if the error management capabilities of the end nodes and the intermediate nodes are the same, they can have a competitive communication capacity.

Since one hop is assumed to be 20 km from the simulation assumption, the 256 nodes treated in this study are approximately 5000 km.
The end-to-end architecture proposed in this study is expected to work at least at the continental level.

Breakthroughs have repeatedly improved the error probabilities of qubits.
By continuously improving error probabilities, the distance and number of hops that can be processed by end-to-end purification will increase.
The quantum Internet should be designed to operate with high efficiency even after the error probability improves.
In this study, we found that even improvement of the error probabilities of end nodes can improve the resource efficiency of the network side.
The quantum Internet may be made even more efficient by using end-to-end properties in other areas besides error management.

This study has shown that it works, but the fidelity is likely to deteriorate as the distance increases in practice.
Therefore, it is essential to take advantage of locality in the quantum Internet, as seen in the effectiveness of CDNs in the current Internet.
To utilize end-to-end protocols and ultra-long distance protocols using side boxes together may be helpful to exploit locality.

Using various technologies rather than a single technology is essential to make the quantum Internet operationally scalable and robust infrastructure.
Less dependence on each other in the processing of communication nodes contributes to compatibility, interoperability, and independently distributed cooperation.
Meanwhile, if the network does not have end-to-end properties, it may become a significant obstacle when introducing new technologies.
In addition to error management, it is essential to actively consider the end-to-end nature of the quantum Internet in future research.

In this study, we have conducted a linear network simulation to investigate an end-to-end connection.
As a future work, the case where multiple communications exist in a network where contention occurs, such as a dumbbell network, should be good to investigate.

\section*{Acknowledgments}

SN acknowledge Kentaro Teramoto, Toshihiko Sasaki, Rodney Van Meter and Yoshihiro Kawahara 
for helpful discussions and comments on an early draft of this manuscript.
SN also acknowledge members of Quantum Internet Task Force, which is a research consortium to realize the Quantum Internet, for comprehensive and interdisciplinary discussions of the Quantum Internet.

\bibliographystyle{plain}
\bibliography{library}

\appendix
\clearpage
\onecolumn

\section{More simulation results}
\label{appendix:mr}

\subsection{Results where error rates in end node are $0.0001$ and error rates in intermediate nodes are $0.0001$, too.}
\begin{figure*}[h]
    \begin{center}
        \includegraphics[width=8.5cm]{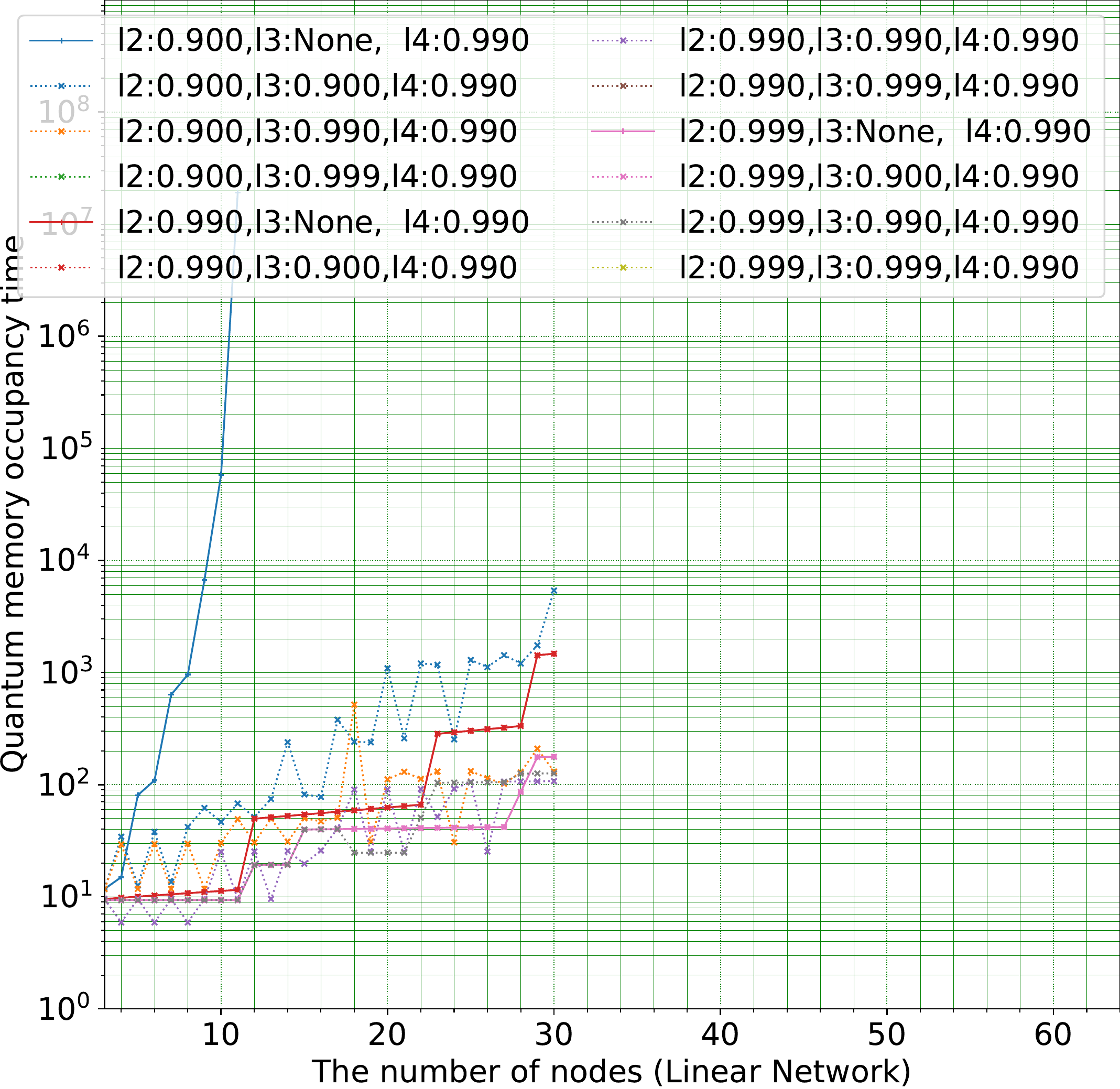}
        \hspace{3mm}
        \includegraphics[width=8.5cm]{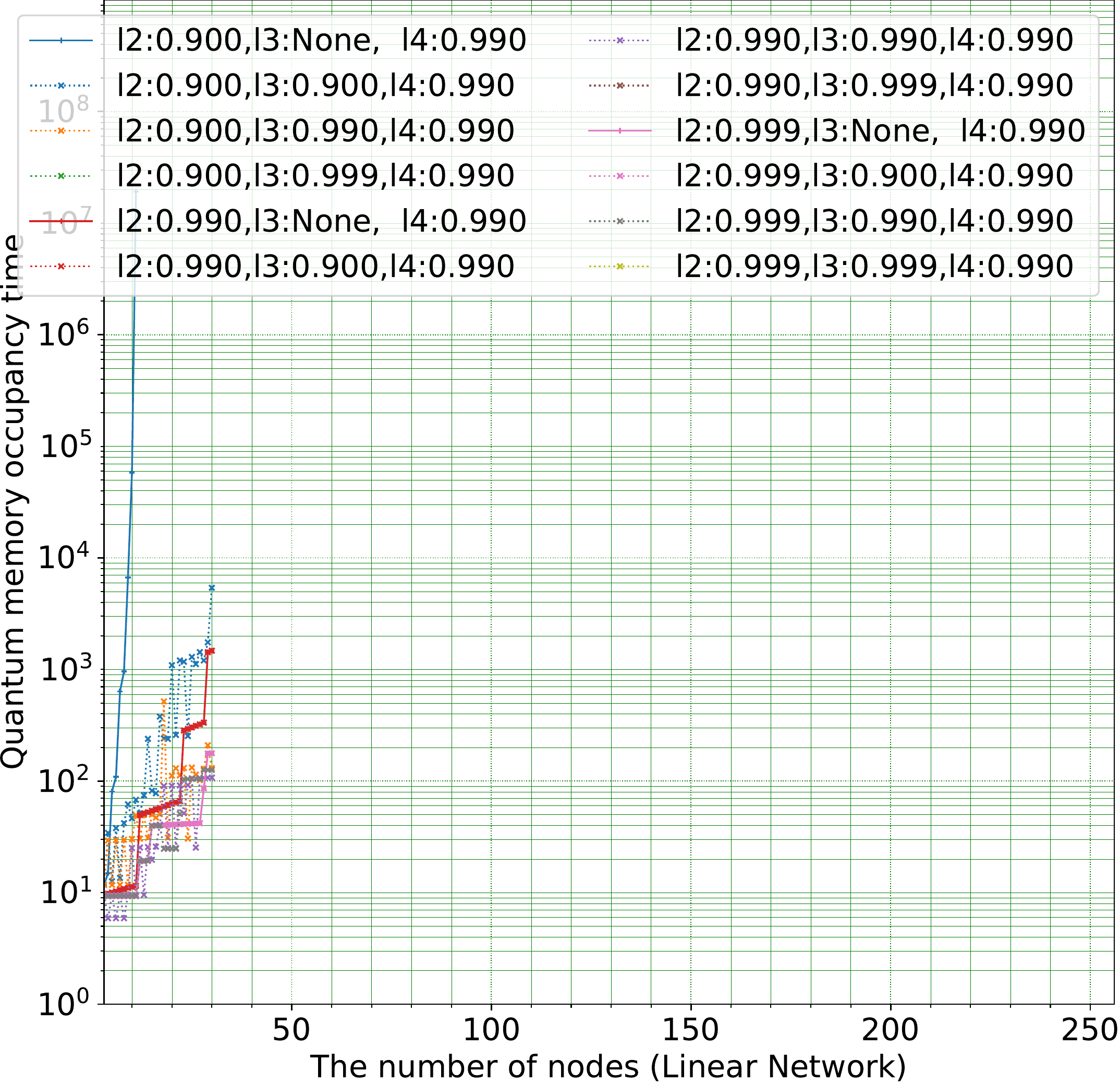}
    \end{center}
    \caption{Quantum memory occupancy time of end nodes where error rates in end node are $0.0001$ and error rates in intermediate nodes are $0.0001$, too.}
    \label{fig:end_4_4_4_4}
  \end{figure*}

\begin{figure*}[h]
    \begin{center}
        \includegraphics[width=8.5cm]{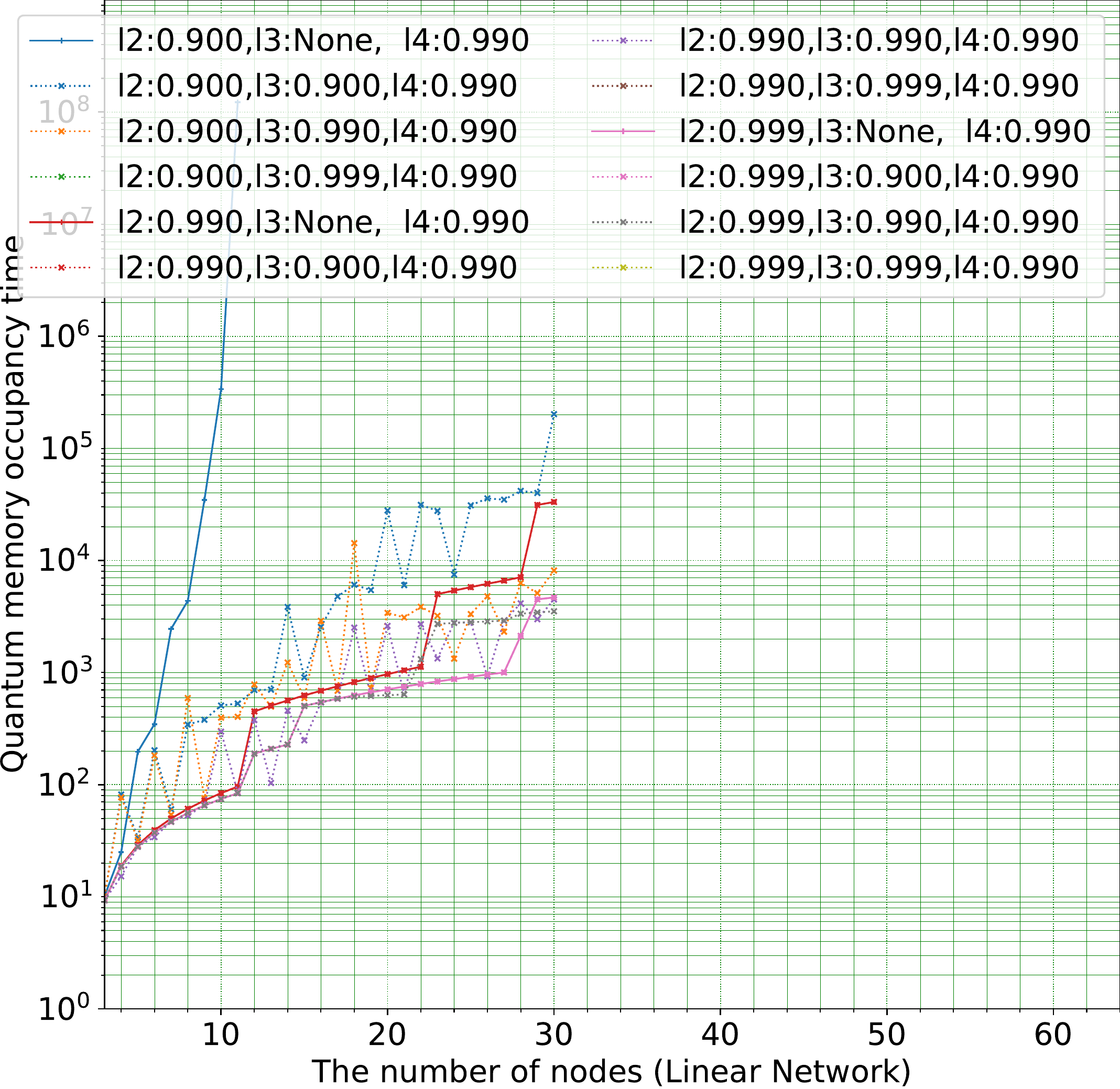}
        \hspace{3mm}
        \includegraphics[width=8.5cm]{figures/Int_4_4_4_4_3_256.pdf}
    \end{center}
    \caption{Quantum memory occupancy time of intermediate nodes where error rates in end node are $0.0001$ and error rates in intermediate nodes are $0.0001$, too.}
    \label{fig:int_4_4_4_4}
  \end{figure*}

\begin{figure*}[h]
    \begin{center}
        \includegraphics[width=8.5cm]{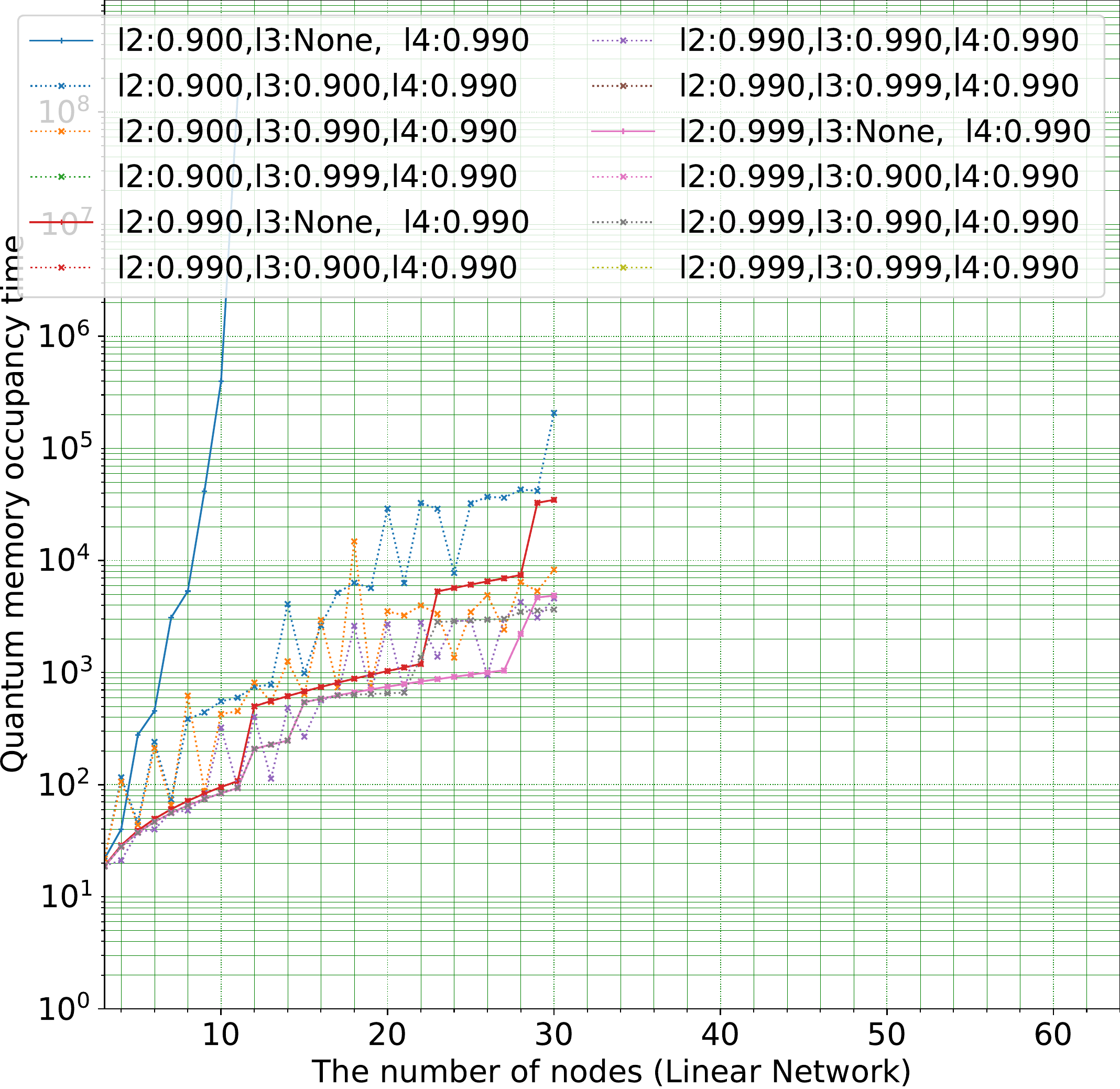}
        \hspace{3mm}
        \includegraphics[width=8.5cm]{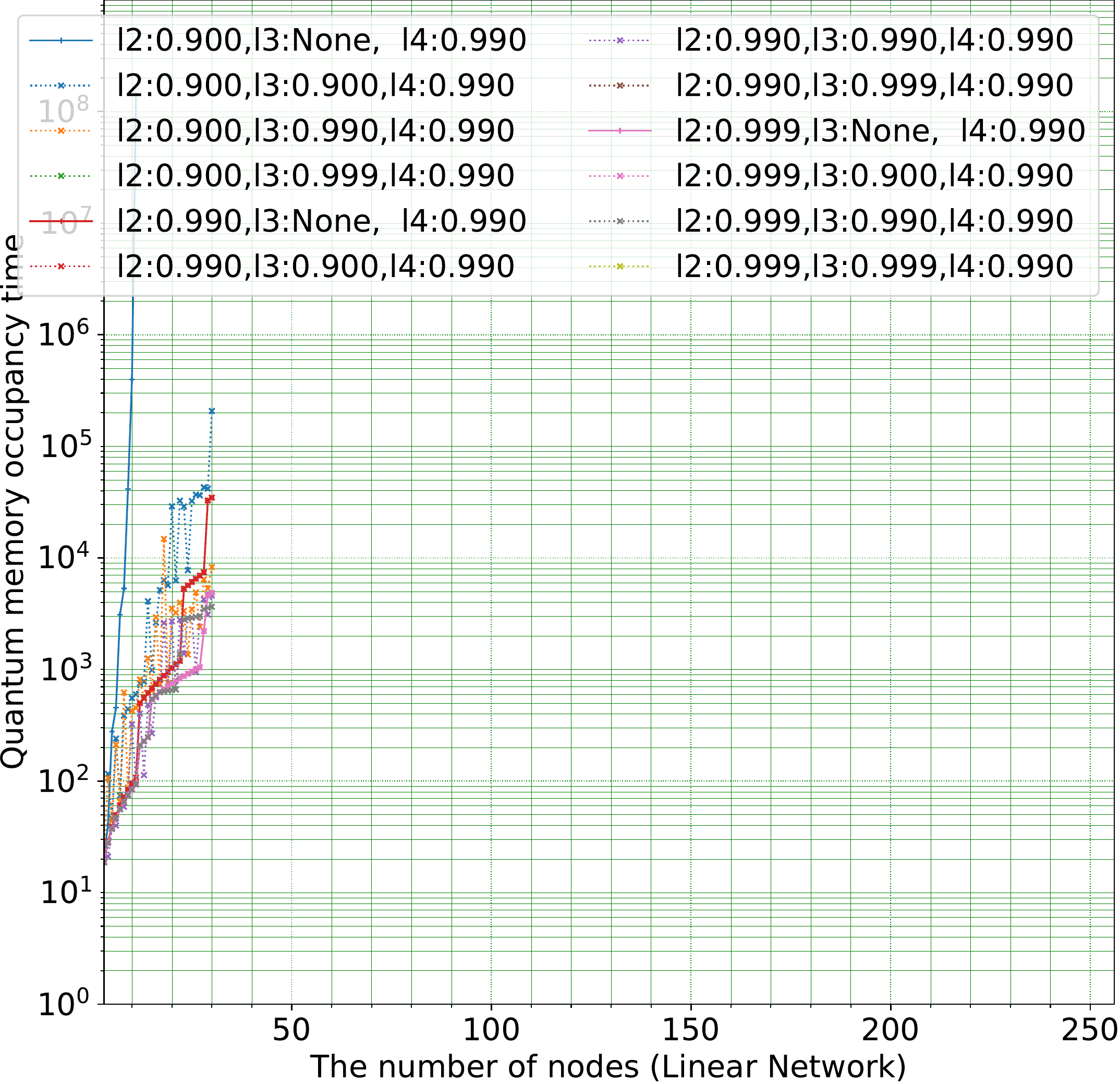}
    \end{center}
    \caption{The total quantum memory occupancy time of where error rates in end node are $0.0001$ and error rates in intermediate nodes are $0.0001$, too.}
    \label{fig:total_4_4_4_4}
  \end{figure*}

Fig. ~\ref{fig:end_4_4_4_4}, ~\ref{fig:int_4_4_4_4} and ~\ref{fig:total_4_4_4_4}
shows the quantum memory occupancy time of end nodes, intermediate nodes, and the total, respectively.

\subsection{Results where error rates in end node are $0.0001$ and error rates in intermediate nodes are $0.00001$.}
\begin{figure*}[h]
  \begin{center}
      \includegraphics[width=8.5cm]{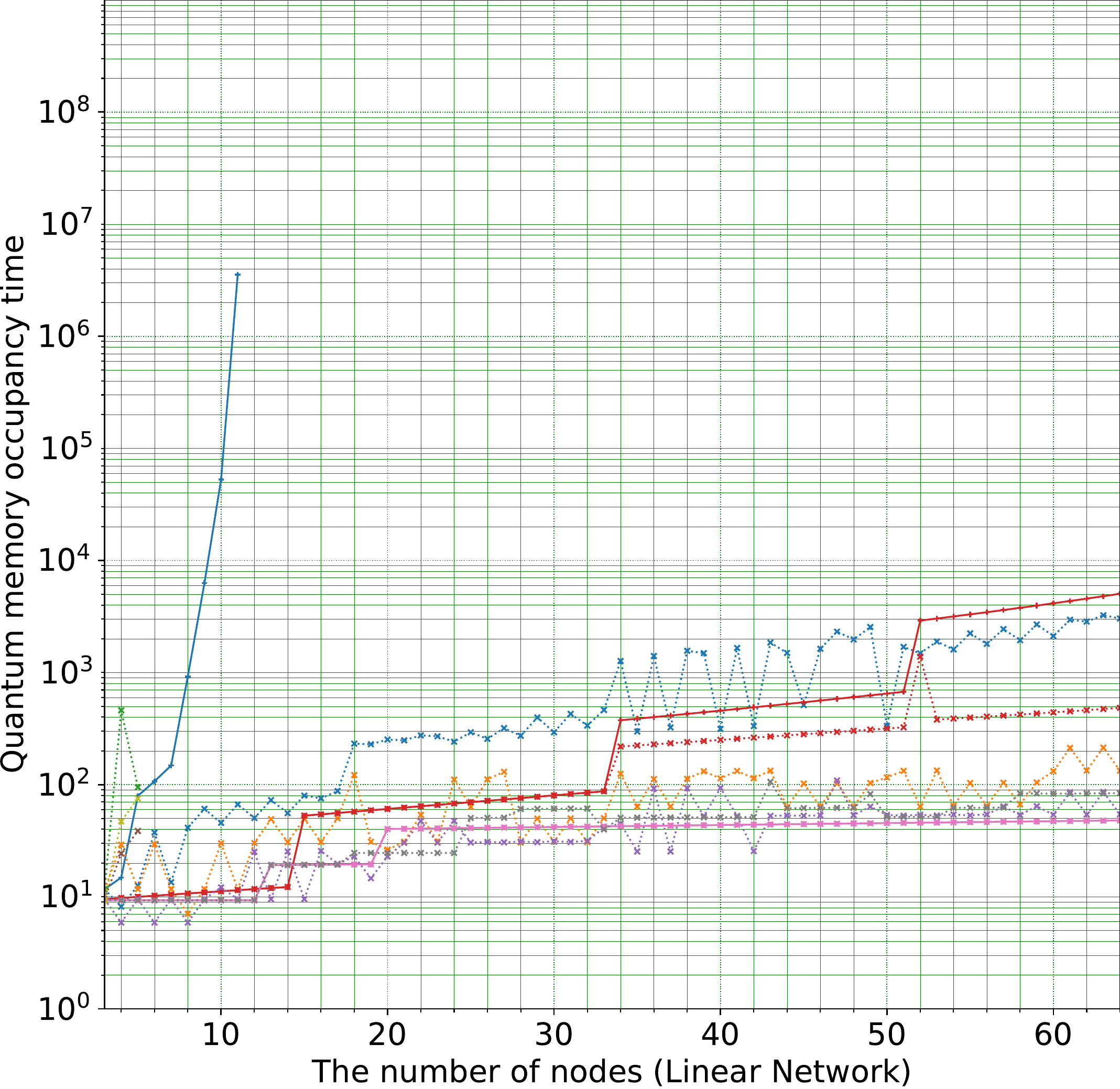}
      \hspace{3mm}
      \includegraphics[width=8.5cm]{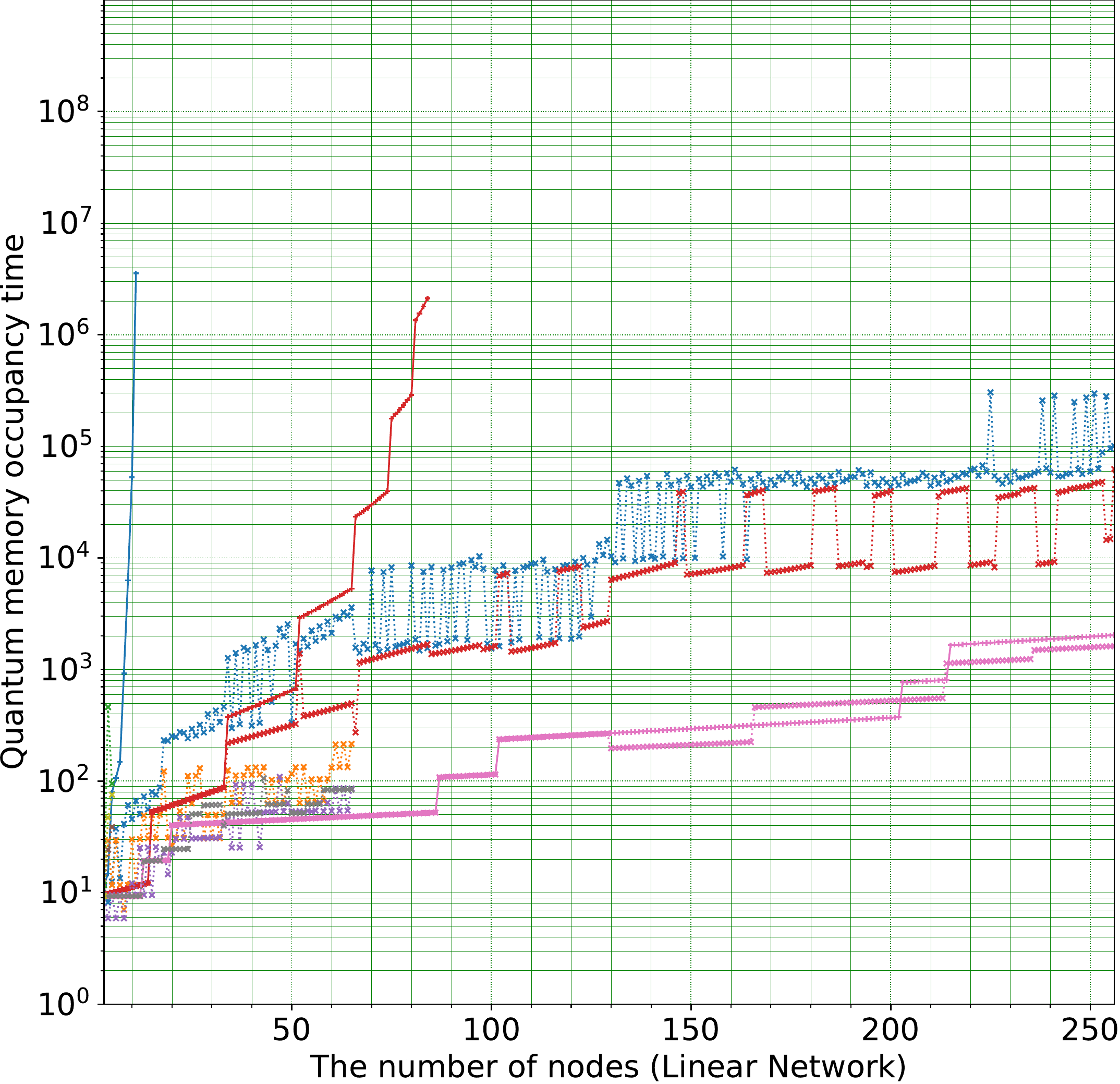}
  \end{center}
  \caption{Quantum memory occupancy time of end nodes where error rates in end node are $0.0001$ and error rates in intermediate nodes are $0.00001$.}
  \label{fig:end_4_4_5_5}
\end{figure*}

\begin{figure*}[h]
  \begin{center}
      \includegraphics[width=8.5cm]{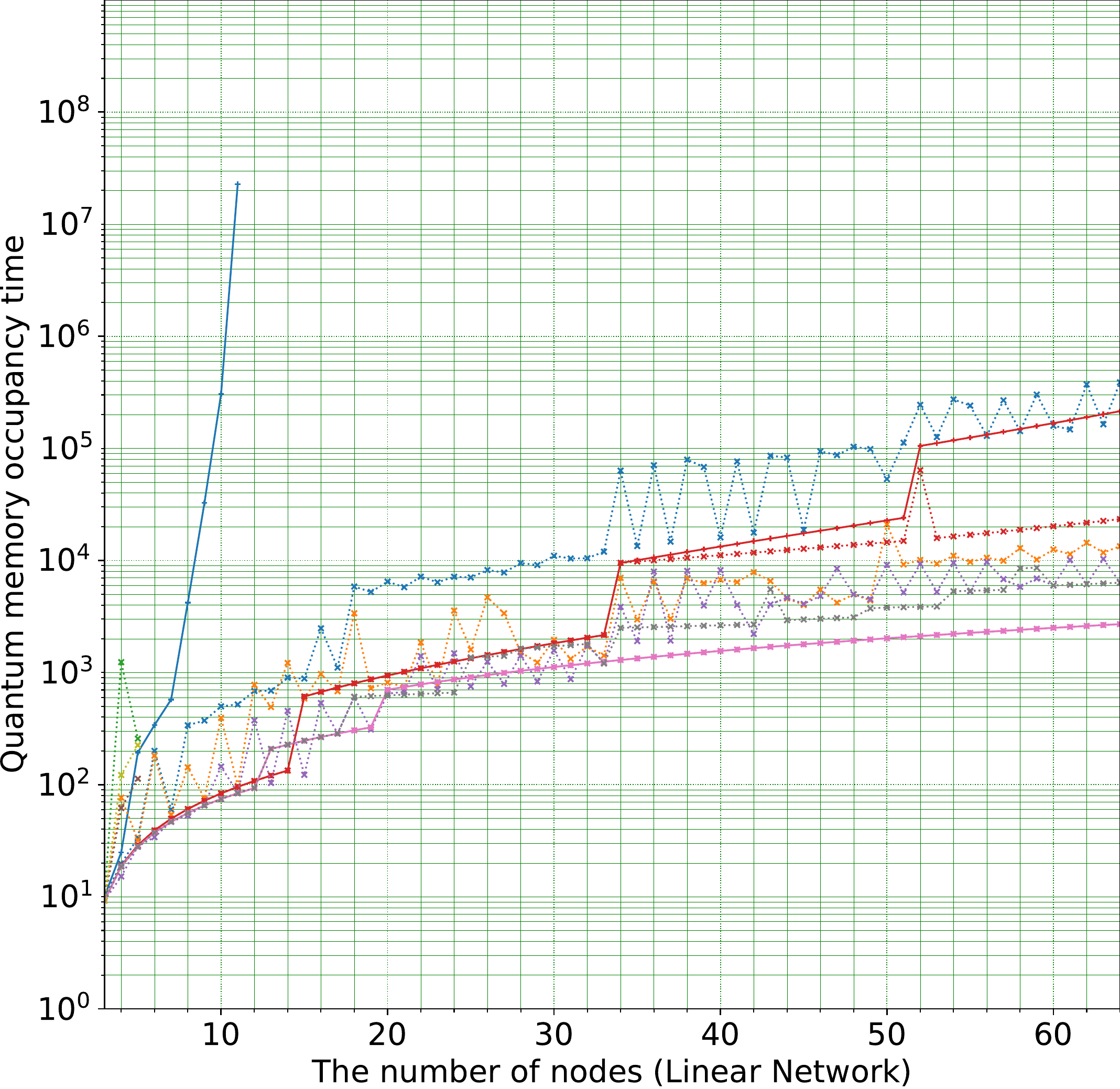}
      \hspace{3mm}
      \includegraphics[width=8.5cm]{figures/Int_4_4_5_5_3_256.pdf}
  \end{center}
  \caption{Quantum memory occupancy time of intermediate nodes where error rates in end node are $0.0001$ and error rates in intermediate nodes are $0.00001$.}
  \label{fig:int_4_4_5_5}
\end{figure*}

\begin{figure*}[h]
    \begin{center}
        \includegraphics[width=8.5cm]{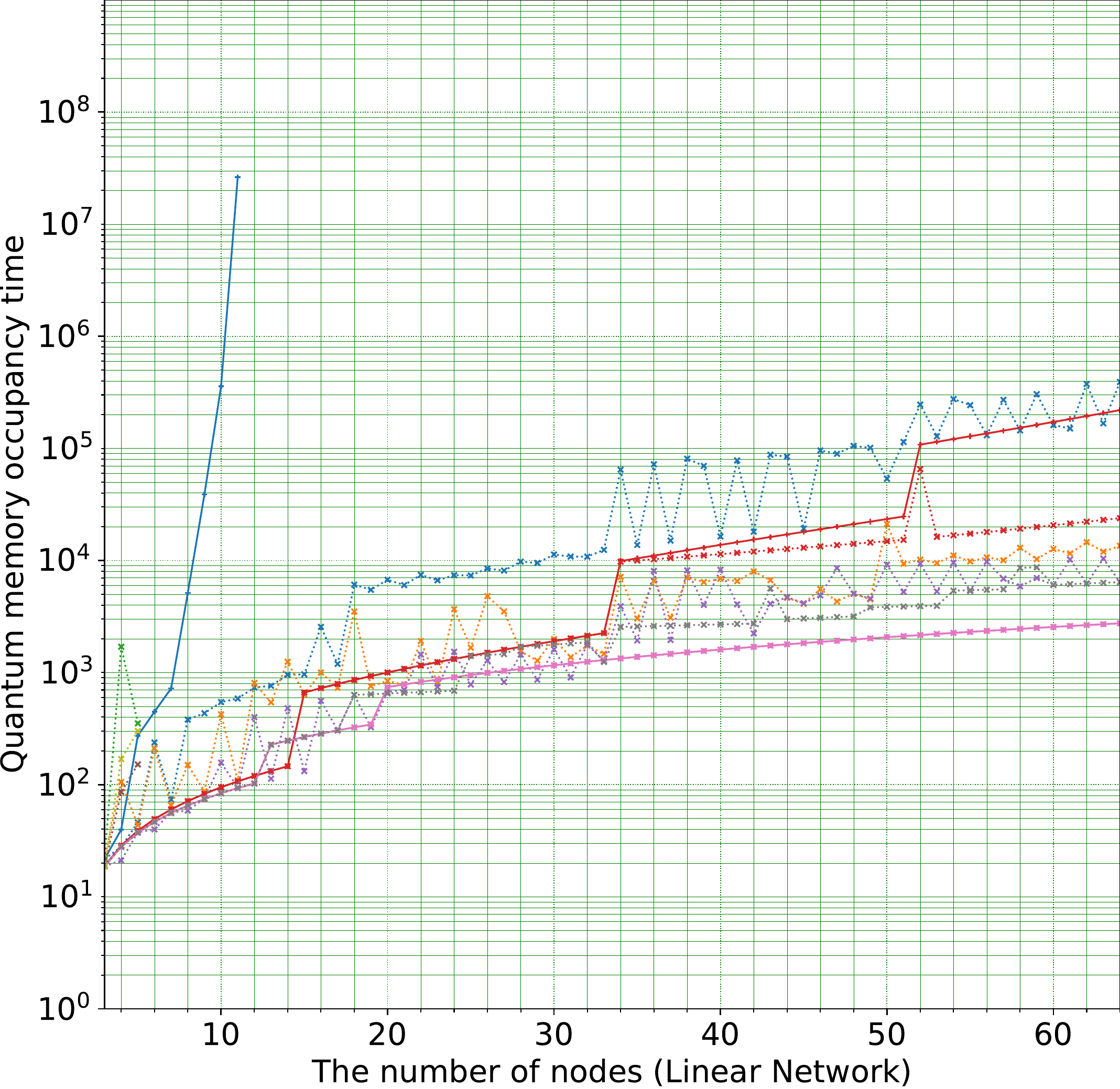}
        \hspace{3mm}
        \includegraphics[width=8.5cm]{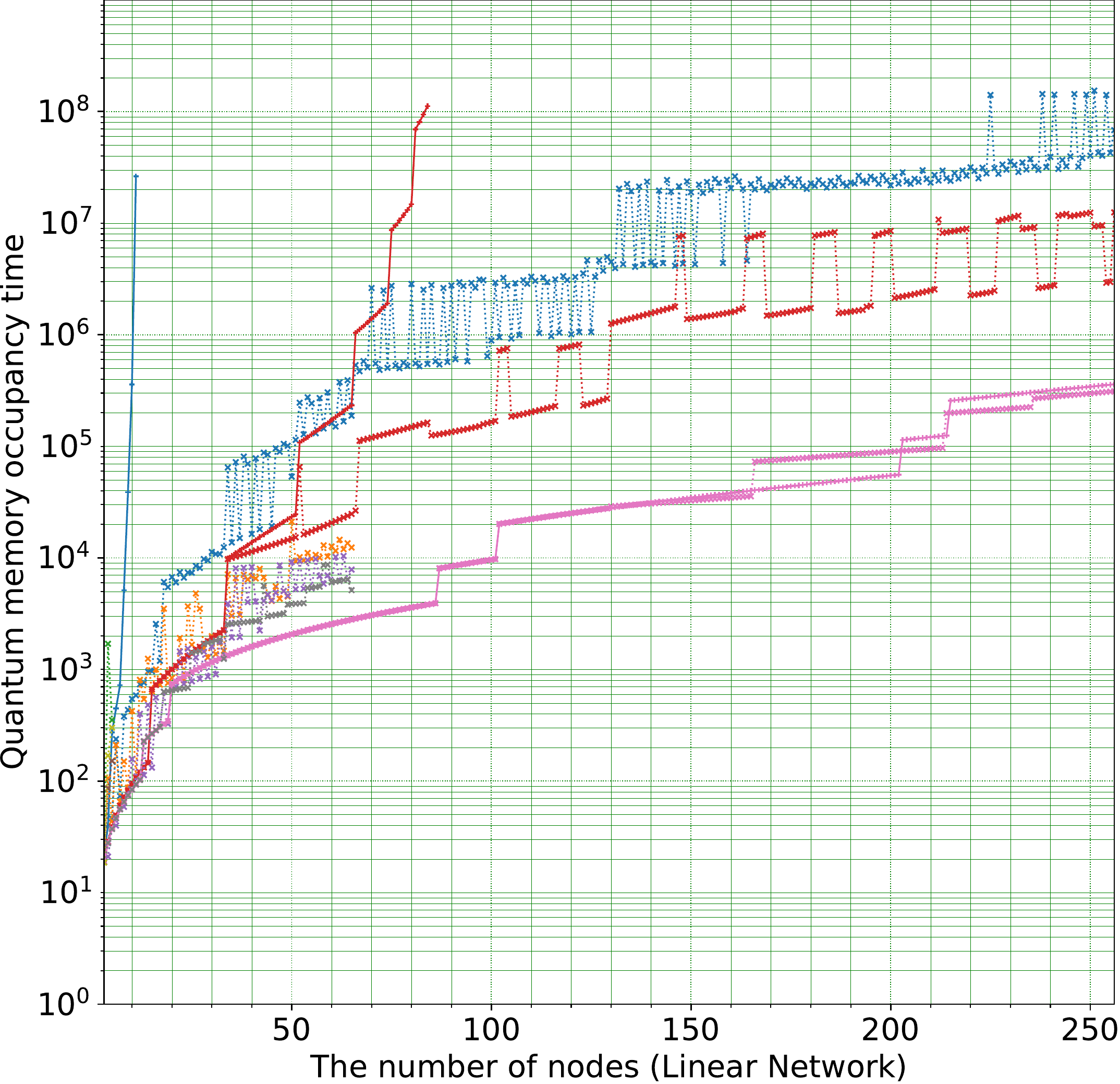}
    \end{center}
    \caption{The total quantum memory occupancy time of where error rates in end node are $0.0001$ and error rates in intermediate nodes are $0.00001$.}
    \label{fig:total_4_4_5_5}
  \end{figure*}

Fig. ~\ref{fig:end_4_4_5_5}, ~\ref{fig:int_4_4_5_5} and ~\ref{fig:total_4_4_5_5}
shows the quantum memory occupancy time of end nodes, intermediate nodes, and the total, respectively.

\subsection{Results where error rates in end node are $0.0001$ and error rates in intermediate nodes are $0$.}
\begin{figure*}[h]
  \begin{center}
      \includegraphics[width=8.5cm]{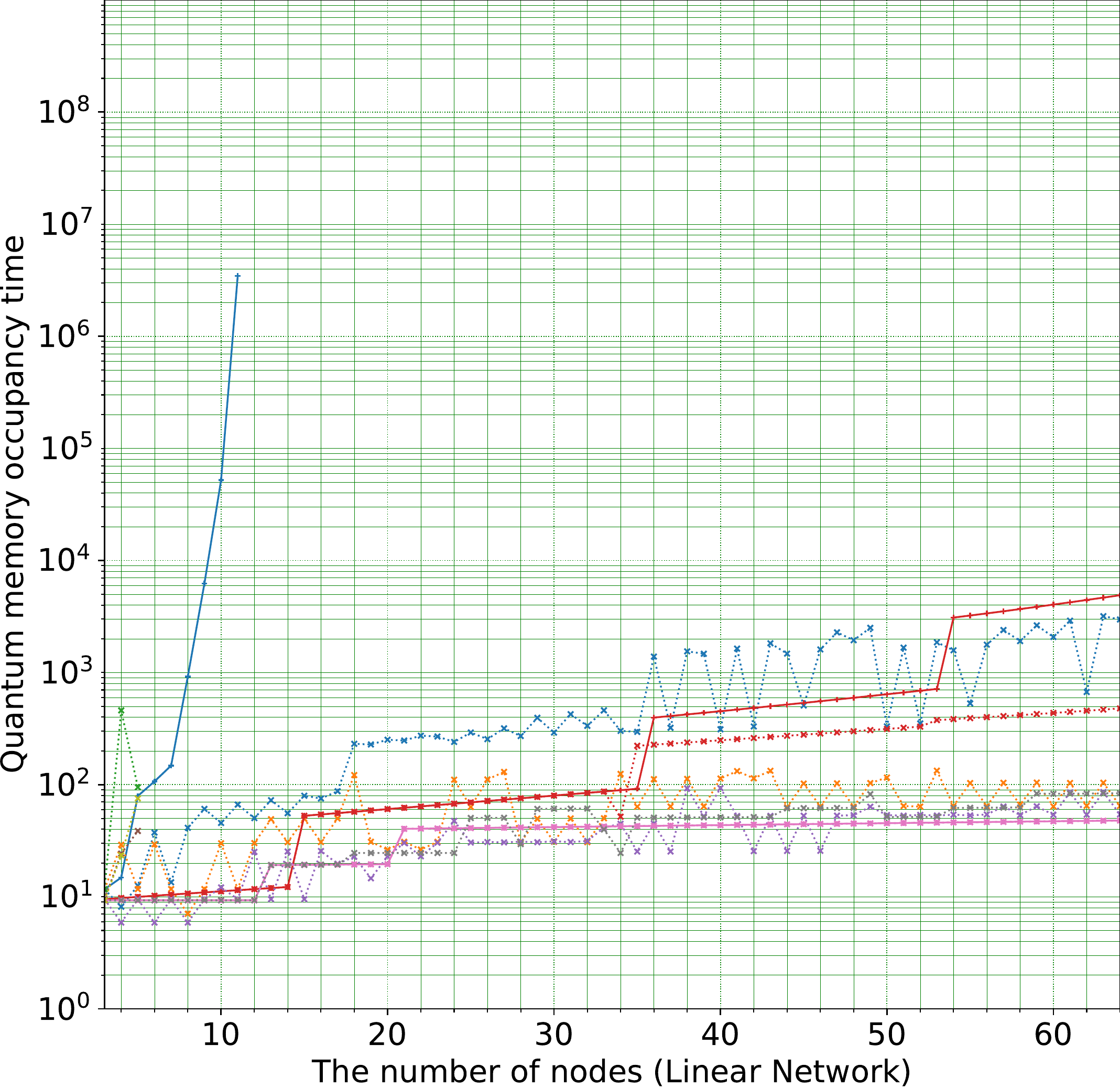}
      \hspace{3mm}
      \includegraphics[width=8.5cm]{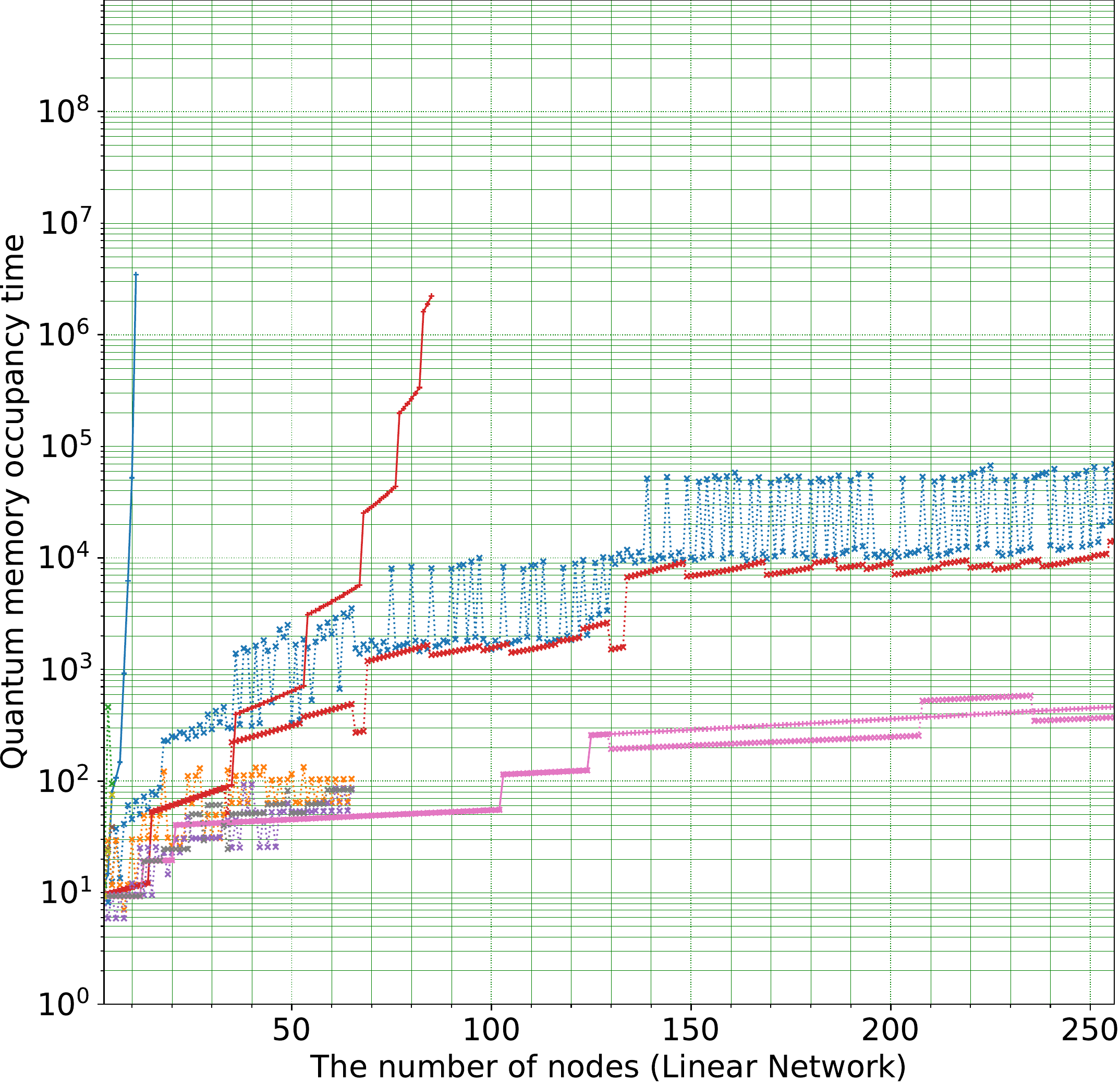}
  \end{center}
  \caption{Quantum memory occupancy time of end nodes where error rates in end node are $0.0001$ and error rates in intermediate nodes are $0$.}
  \label{fig:end_4_4_0_0}
\end{figure*}

\begin{figure*}[h]
  \begin{center}
      \includegraphics[width=8.5cm]{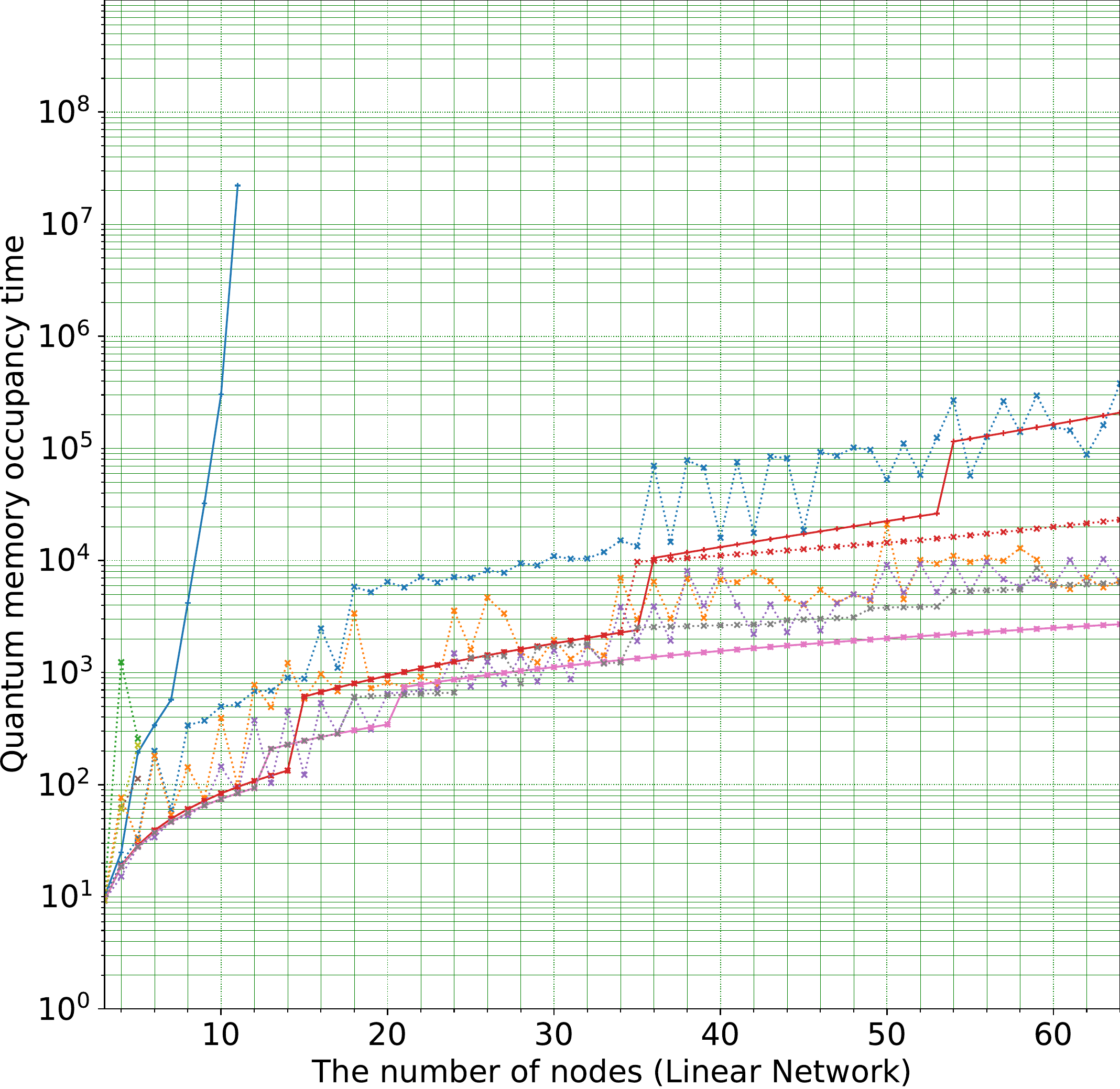}
      \hspace{3mm}
      \includegraphics[width=8.5cm]{figures/Int_4_4_0_0_3_256.pdf}
  \end{center}
  \caption{Quantum memory occupancy time of intermediate nodes where error rates in end node are $0.0001$ and error rates in intermediate nodes are $0$.}
  \label{fig:int_4_4_0_0}
\end{figure*}

\begin{figure*}[h]
    \begin{center}
        \includegraphics[width=8.5cm]{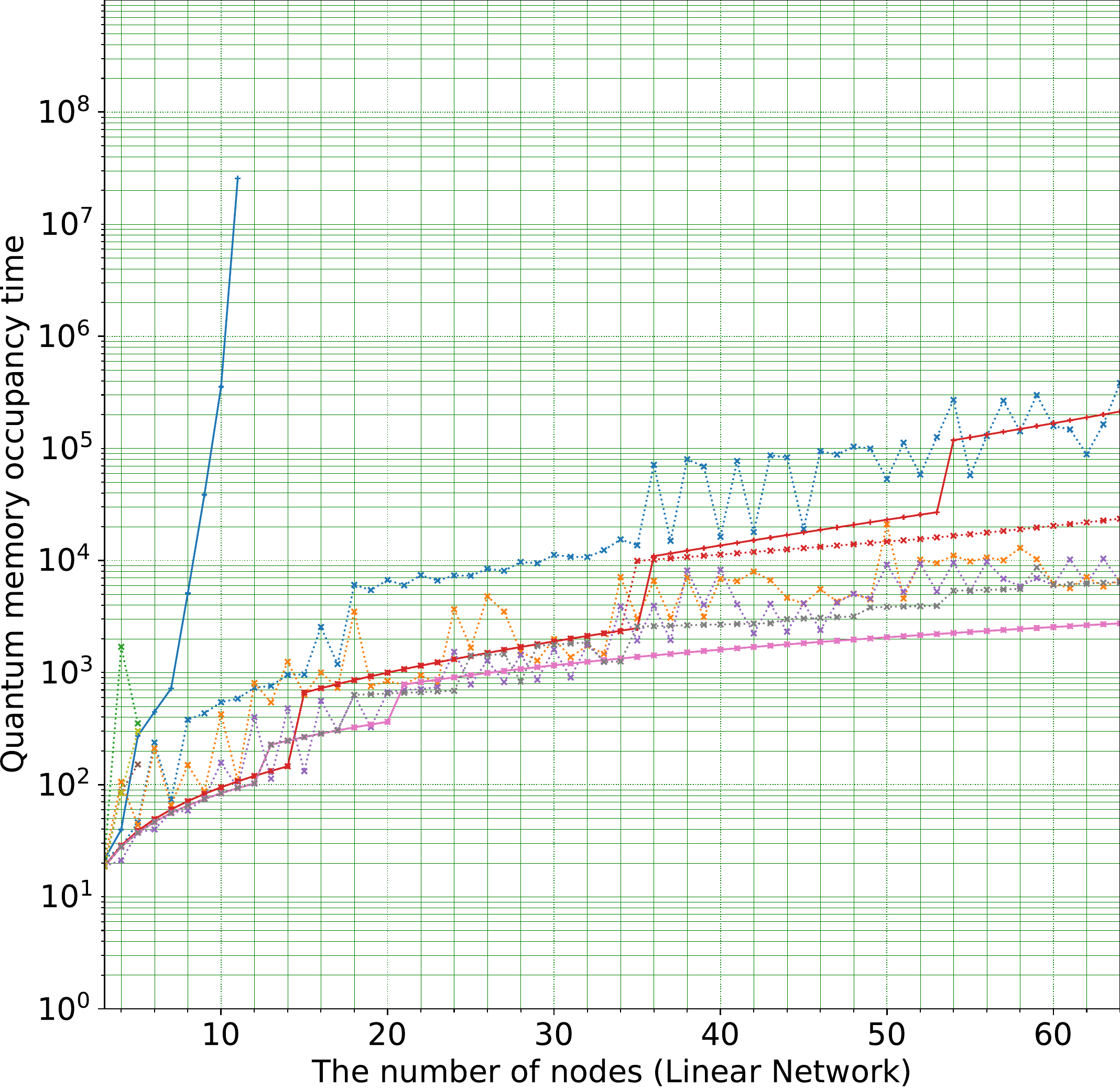}
        \hspace{3mm}
        \includegraphics[width=8.5cm]{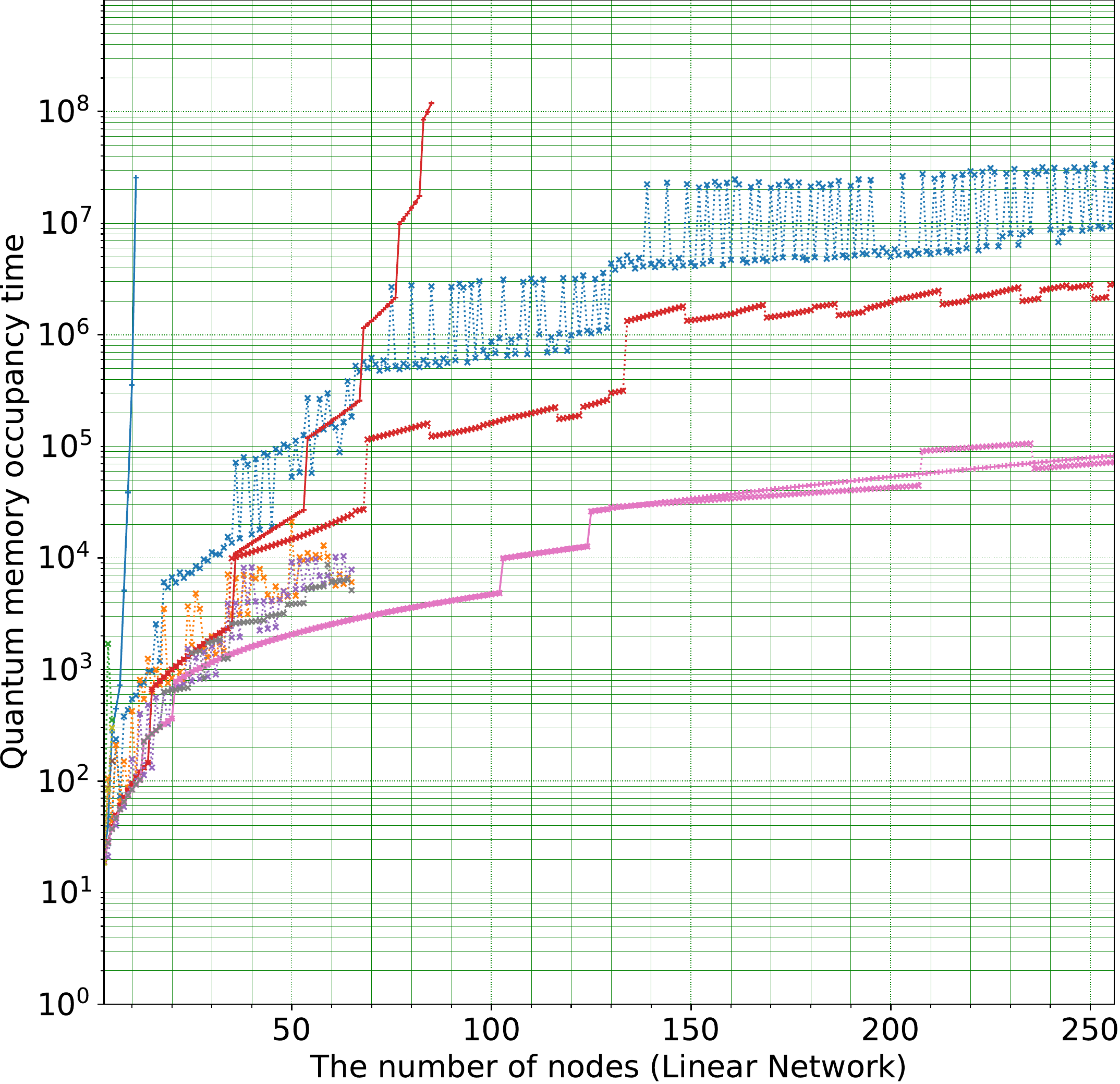}
    \end{center}
    \caption{The total quantum memory occupancy time of where error rates in end node are $0.0001$ and error rates in intermediate nodes are $0$.}
    \label{fig:total_4_4_0_0}
  \end{figure*}

Fig. ~\ref{fig:end_4_4_0_0}, ~\ref{fig:int_4_4_0_0} and ~\ref{fig:total_4_4_0_0}
shows the quantum memory occupancy time of end nodes, intermediate nodes, and the total, respectively.

  \subsection{Results where error rates in end node are $0.00001$ and error rates in intermediate nodes are $0.00001$.}
\begin{figure*}[h]
    \begin{center}
        \includegraphics[width=8.5cm]{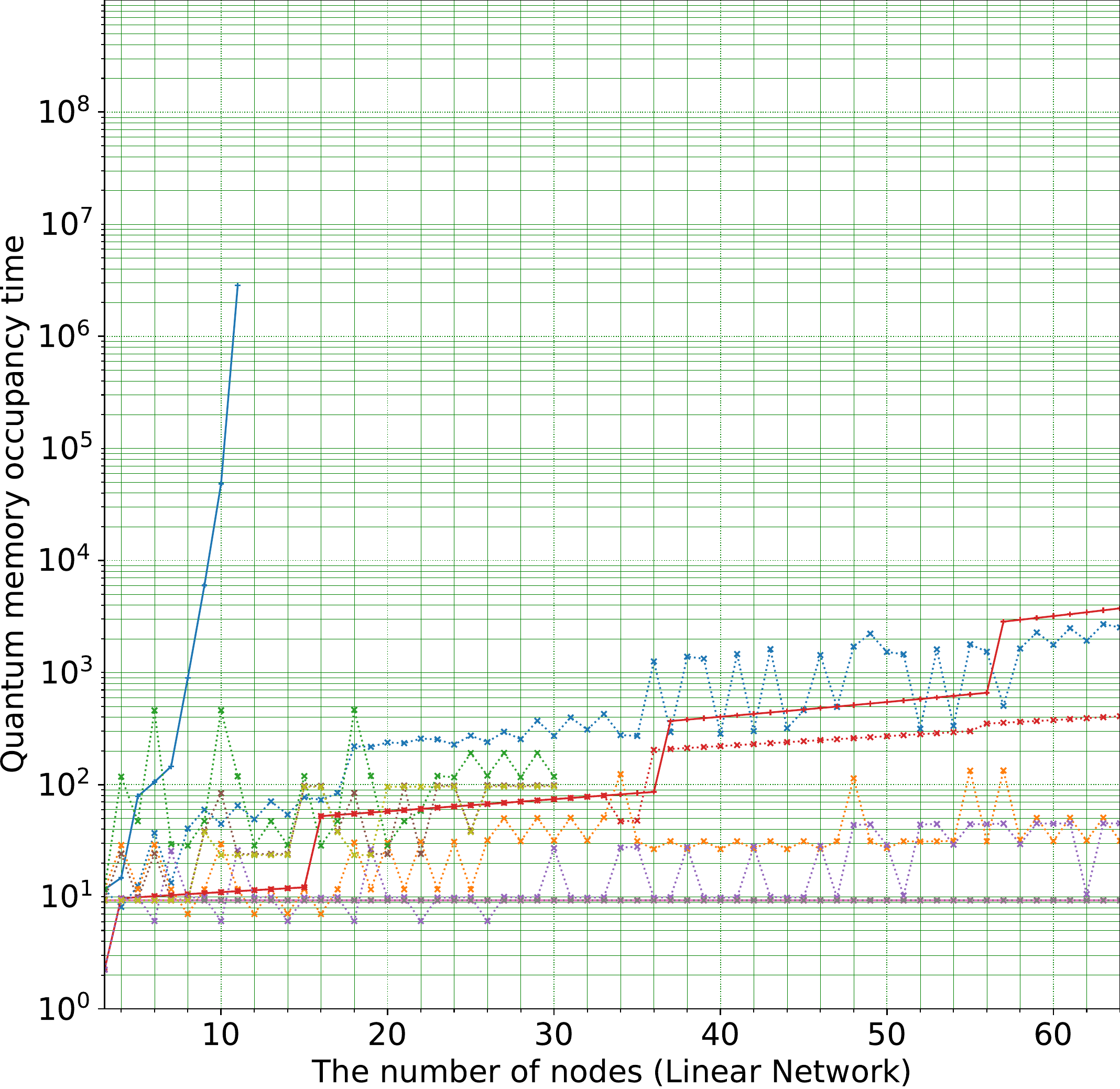}
        \hspace{3mm}
        \includegraphics[width=8.5cm]{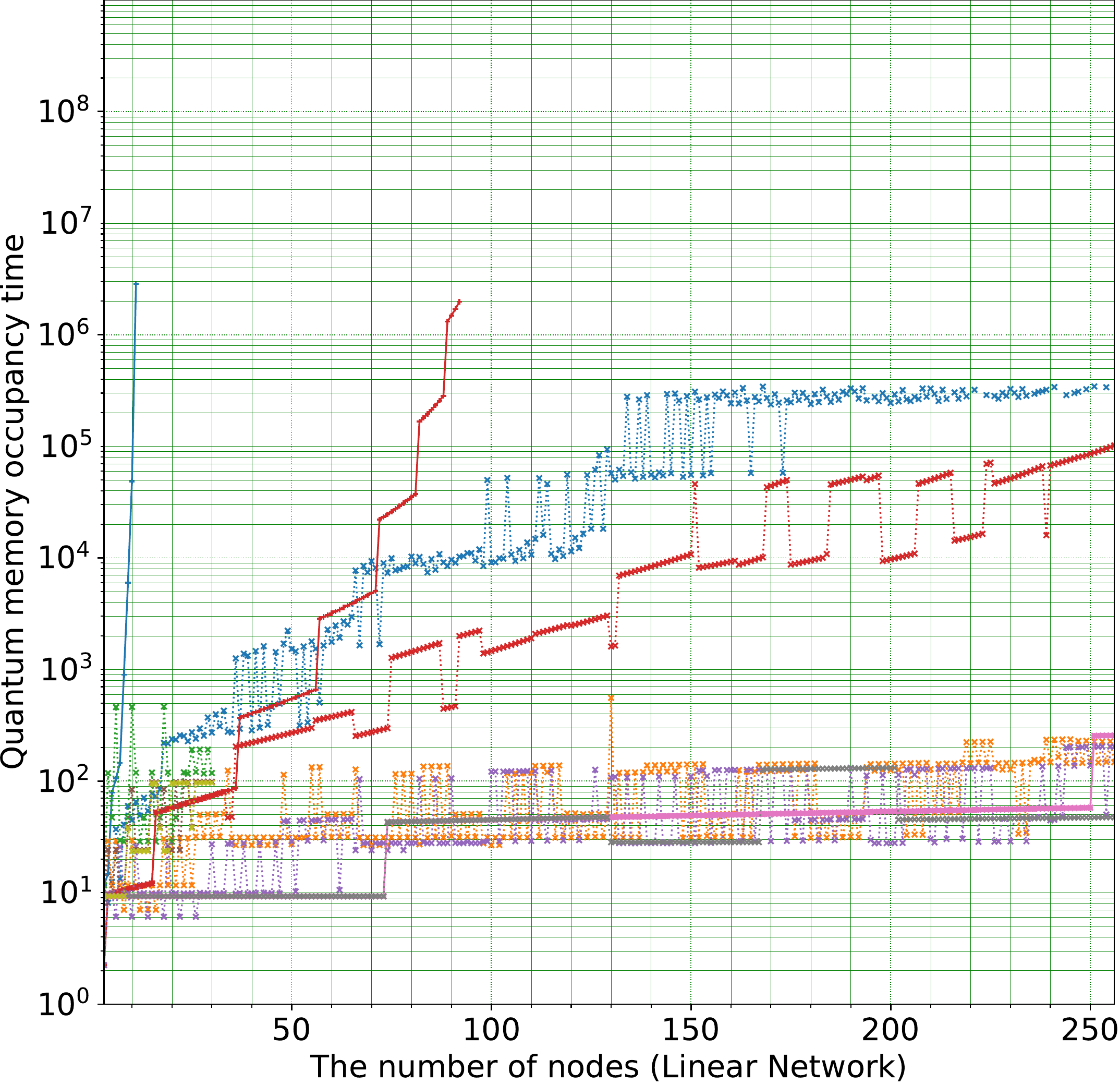}
    \end{center}
    \caption{Quantum memory occupancy time of end nodes where error rates in end node are $0.00001$ and error rates in intermediate nodes are $0.00001$.}
    \label{fig:end_5_5_5_5}
  \end{figure*}

\begin{figure*}[h]
    \begin{center}
        \includegraphics[width=8.5cm]{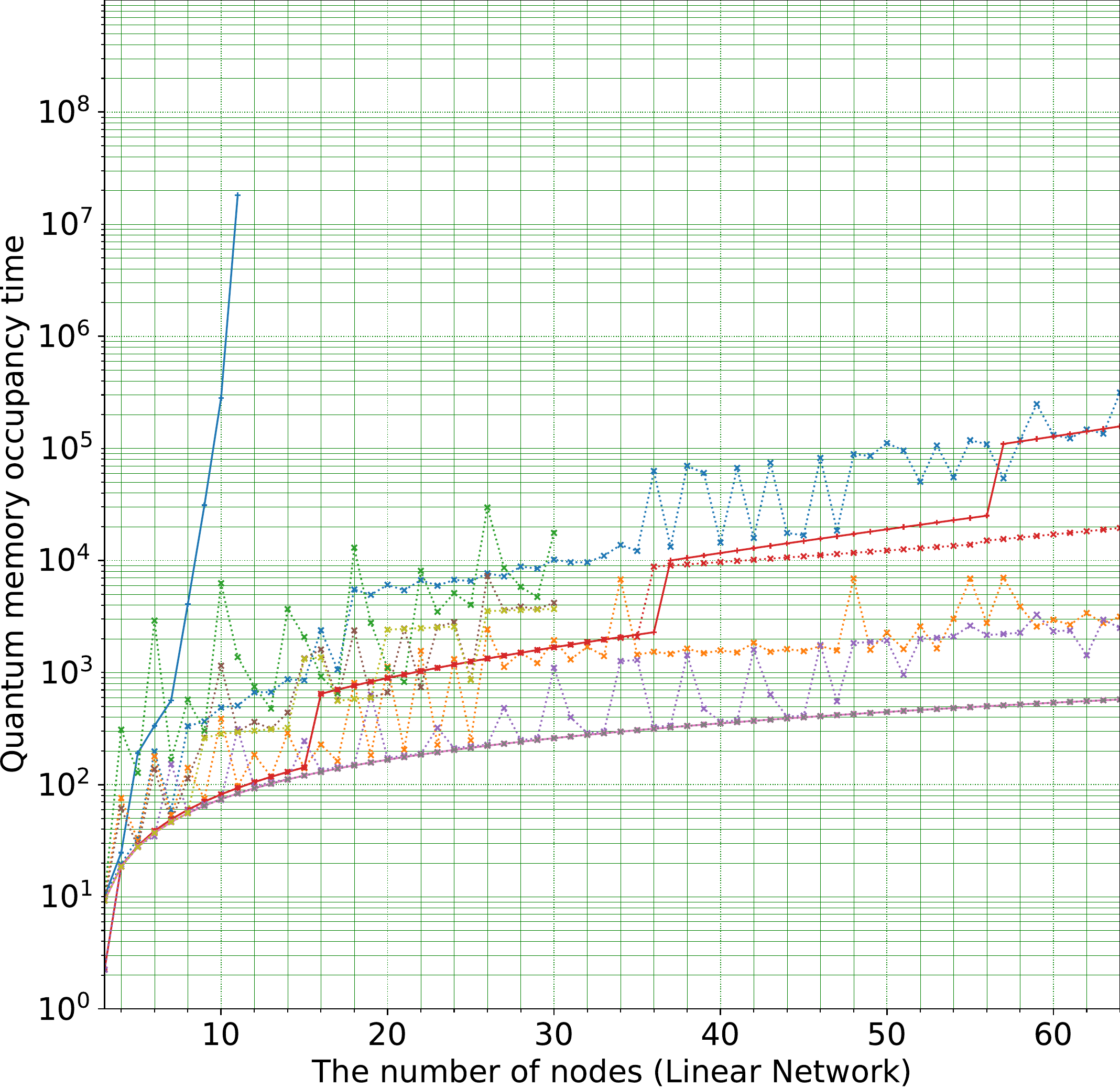}
        \hspace{3mm}
        \includegraphics[width=8.5cm]{figures/Int_5_5_5_5_3_256.pdf}
    \end{center}
    \caption{Quantum memory occupancy time of intermediate nodes where error rates in end node are $0.00001$ and error rates in intermediate nodes are $0.00001$.}
    \label{fig:int_5_5_5_5}
  \end{figure*}

\begin{figure*}[h]
    \begin{center}
        \includegraphics[width=8.5cm]{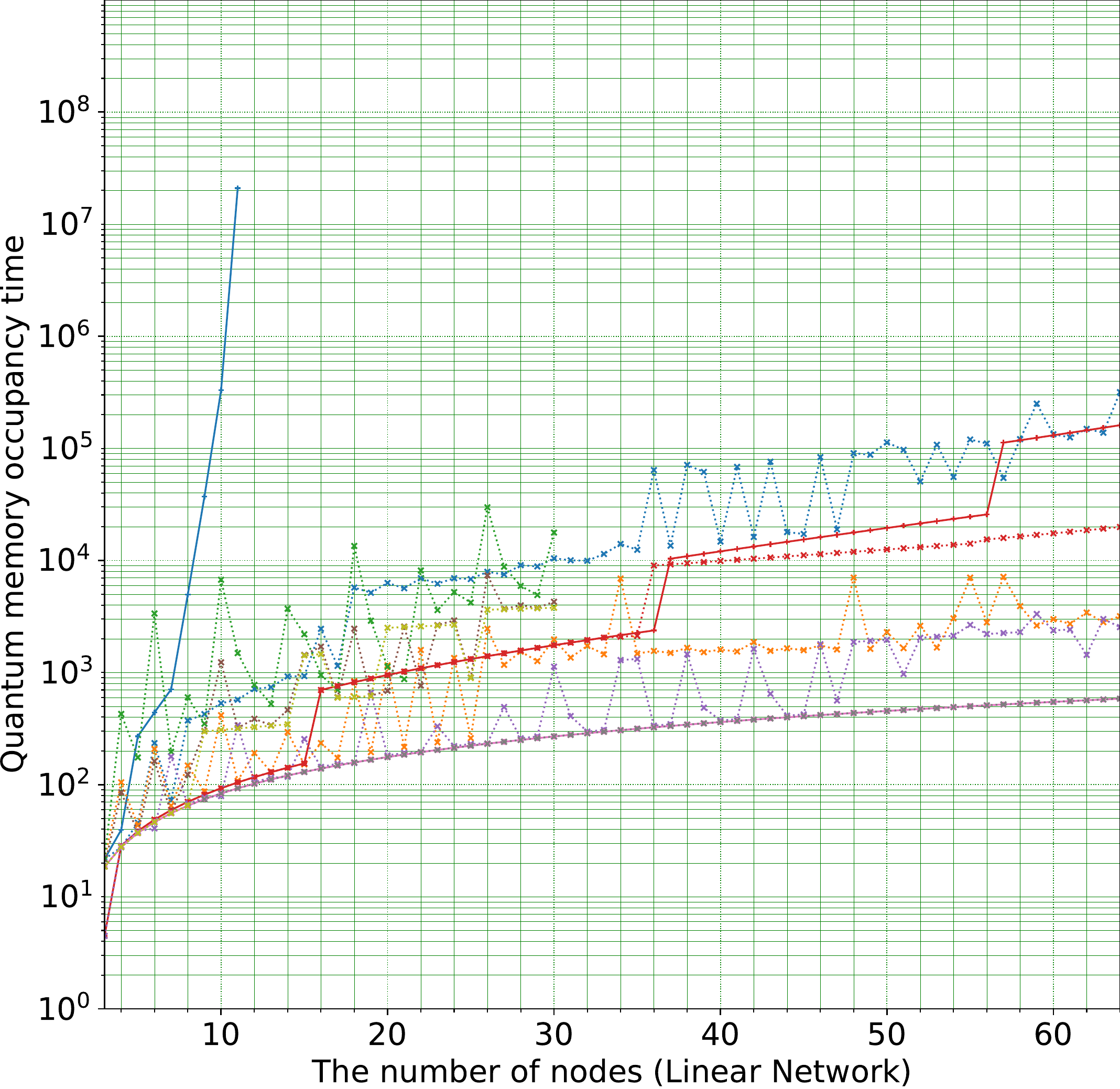}
        \hspace{3mm}
        \includegraphics[width=8.5cm]{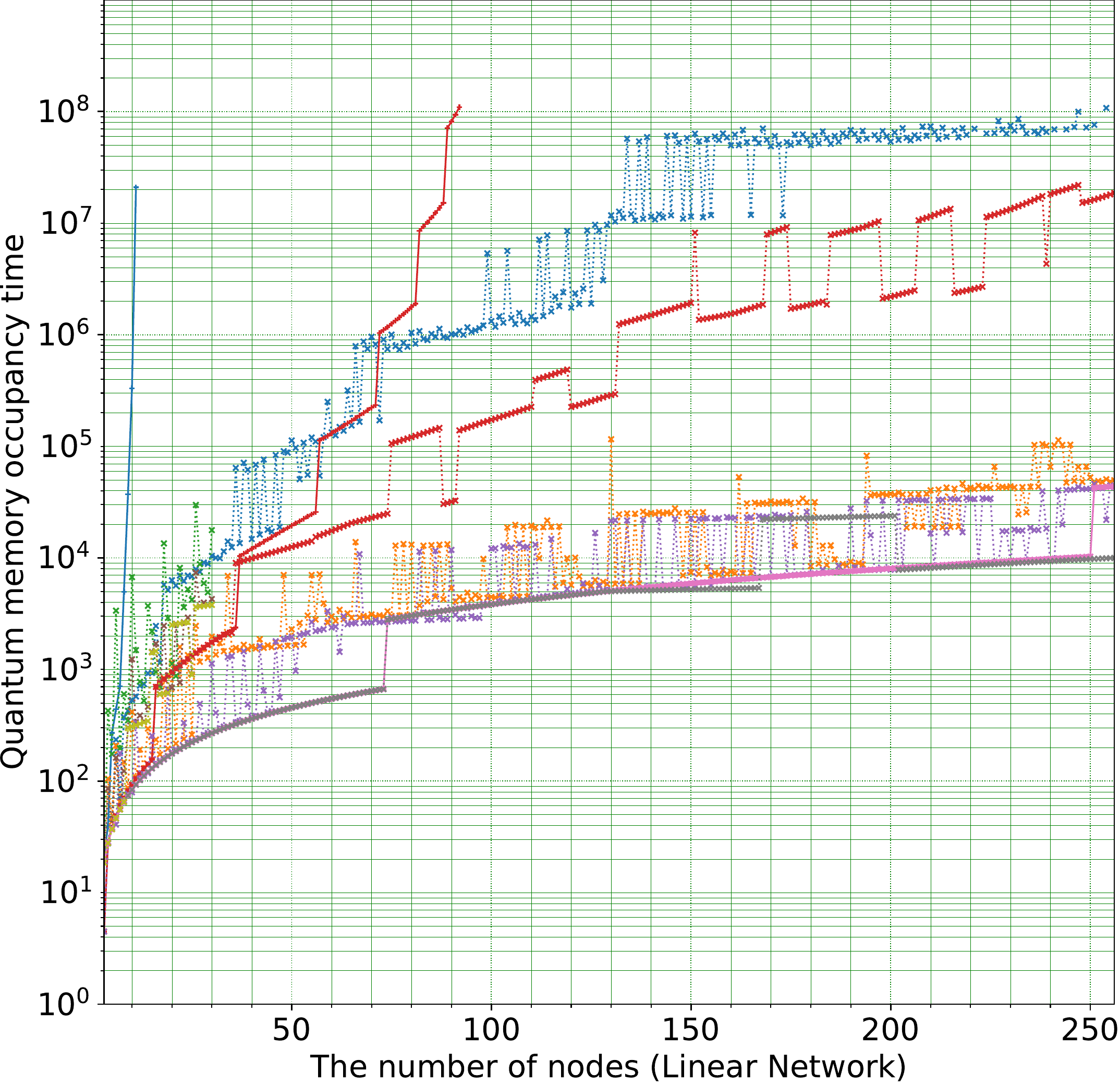}
    \end{center}
    \caption{The total quantum memory occupancy time of where error rates in end node are $0.00001$ and error rates in intermediate nodes are $0.00001$.}
    \label{fig:total_5_5_5_5}
  \end{figure*}

  Fig. ~\ref{fig:end_5_5_5_5}, ~\ref{fig:int_5_5_5_5} and ~\ref{fig:total_5_5_5_5}
  shows the quantum memory occupancy time of end nodes, intermediate nodes, and the total, respectively.
  
\end{document}